\documentclass[12pt]{article}
\usepackage[dvips]{graphicx}
\usepackage{epsfig}
\usepackage{latexsym,amsmath,amsfonts,amssymb}
\usepackage[latin1]{inputenc}
\usepackage[american]{babel}
\usepackage[dvips]{graphicx}
\usepackage{bbm}
\def\im{Invent. Math.}
\def\atmp{Adv. Theor. Math. Phys. }

\def\atmp{Adv. Theor. Math. Phys.}

\newcommand{\be}{\begin{equation}}
\newcommand{\ee}{\end{equation}}
\newcommand{\beq}{\begin{equation}}
\newcommand{\eeq}{\end{equation}}
\newcommand{\bea}{\begin{eqnarray}}
\newcommand{\eea}{\end{eqnarray}}

\newcommand{\ba}{\begin{eqnarray}}
\newcommand{\ea}{\end{eqnarray}}
\begin{document}
\baselineskip=15.5pt
\pagestyle{plain}
\setcounter{page}{1}


\def\del{{\partial}}
\def\vev#1{\left\langle #1 \right\rangle}
\def\cn{{\cal N}}
\def\co{{\cal O}}
\def\IC{{\mathbb C}}
\def\IR{{\mathbb R}}
\def\IZ{{\mathbb Z}}
\def\RP{{\bf RP}}
\def\CP{{\bf CP}}
\def\Poincare{{Poincar\'e }}
\def\tr{{\rm tr}}
\def\tp{{\tilde \Phi}}

\def\TL{\hfil$\displaystyle{##}$}
\def\TR{$\displaystyle{{}##}$\hfil}
\def\TC{\hfil$\displaystyle{##}$\hfil}
\def\TT{\hbox{##}}
\def\HLINE{\noalign{\vskip1\jot}\hline\noalign{\vskip1\jot}}
\def\seqalign#1#2{\vcenter{\openup1\jot
   \halign{\strut #1\cr #2 \cr}}}
\def\lbldef#1#2{\expandafter\gdef\csname #1\endcsname {#2}}
\def\eqn#1#2{\lbldef{#1}{(\ref{#1})}%
\begin{equation} #2 \label{#1} \end{equation}}
\def\eqalign#1{\vcenter{\openup1\jot
     \halign{\strut\span\TL & \span\TR\cr #1 \cr
    }}}
\def\eno#1{(\ref{#1})}
\def\href#1#2{#2}
\def\half{{1 \over 2}}

\def\ads{{\it AdS}}
\def\adsp{{\it AdS}$_{p+2}$}
\def\cft{{\it CFT}}

\newcommand{\ber}{\begin{eqnarray}}
\newcommand{\eer}{\end{eqnarray}}

\newcommand{\beqar}{\begin{eqnarray}}
\newcommand{\cN}{{\cal N}}
\newcommand{\cO}{{\cal O}}
\newcommand{\cA}{{\cal A}}
\newcommand{\cT}{{\cal T}}
\newcommand{\cF}{{\cal F}}
\newcommand{\cC}{{\cal C}}
\newcommand{\cR}{{\cal R}}
\newcommand{\cW}{{\cal W}}
\newcommand{\eeqar}{\end{eqnarray}}
\newcommand{\tht}{\thteta}
\newcommand{\lm}{\lambda}\newcommand{\Lm}{\Lambda}
\newcommand{\eps}{\epsilon}


\newcommand{\nonu}{\nonumber}
\newcommand{\oh}{\displaystyle{\frac{1}{2}}}
\newcommand{\dsl}
   {\kern.06em\hbox{\raise.15ex\hbox{$/$}\kern-.56em\hbox{$\partial$}}}
\newcommand{\id}{i\!\!\not\!\partial}
\newcommand{\as}{\not\!\! A}
\newcommand{\ps}{\not\! p}
\newcommand{\ks}{\not\! k}
\newcommand{\D}{{\cal{D}}}
\newcommand{\dv}{d^2x}
\newcommand{\Z}{{\cal Z}}
\newcommand{\N}{{\cal N}}
\newcommand{\Dsl}{\not\!\! D}
\newcommand{\Bsl}{\not\!\! B}
\newcommand{\Psl}{\not\!\! P}
\newcommand{\eeqarr}{\end{eqnarray}}
\newcommand{\ZZ}{{\rm \kern 0.275em Z \kern -0.92em Z}\;}


\def\del{{\delta^{\hbox{\sevenrm B}}}} \def\ex{{\hbox{\rm e}}}
\def\azb{A_{\bar z}} \def\az{A_z} \def\bzb{B_{\bar z}} \def\bz{B_z}
\def\czb{C_{\bar z}} \def\cz{C_z} \def\dzb{D_{\bar z}} \def\dz{D_z}
\def\im{{\hbox{\rm Im}}} \def\mod{{\hbox{\rm mod}}} \def\tr{{\hbox{\rm Tr}}}
\def\ch{{\hbox{\rm ch}}} \def\imp{{\hbox{\sevenrm Im}}}
\def\trp{{\hbox{\sevenrm Tr}}} \def\vol{{\hbox{\rm Vol}}}
\def\rl{\Lambda_{\hbox{\sevenrm R}}} \def\wl{\Lambda_{\hbox{\sevenrm W}}}
\def\fc{{\cal F}_{k+\cox}} \def\vev{vacuum expectation value}
\def\nodiv{\mid{\hbox{\hskip-7.8pt/}}}
\def\ie{{\em i.e.}}
\def\ie{\hbox{\it i.e.}}

\def\CC{{\mathchoice
{\rm C\mkern-8mu\vrule height1.45ex depth-.05ex
width.05em\mkern9mu\kern-.05em}
{\rm C\mkern-8mu\vrule height1.45ex depth-.05ex
width.05em\mkern9mu\kern-.05em}
{\rm C\mkern-8mu\vrule height1ex depth-.07ex
width.035em\mkern9mu\kern-.035em}
{\rm C\mkern-8mu\vrule height.65ex depth-.1ex
width.025em\mkern8mu\kern-.025em}}}

\def\RR{{\rm I\kern-1.6pt {\rm R}}}
\def\NN{{\rm I\!N}}
\def\ZZ{{\rm Z}\kern-3.8pt {\rm Z} \kern2pt}
\def\IB{\relax{\rm I\kern-.18em B}}
\def\ID{\relax{\rm I\kern-.18em D}}
\def\II{\relax{\rm I\kern-.18em I}}
\def\IP{\relax{\rm I\kern-.18em P}}
\newcommand{\CS}{{\scriptstyle {\rm CS}}}
\newcommand{\CSs}{{\scriptscriptstyle {\rm CS}}}
\newcommand{\rc}{\nonumber\\}
\newcommand{\bear}{\begin{eqnarray}}
\newcommand{\eear}{\end{eqnarray}}
\newcommand{\W}{{\cal W}}
\newcommand{\F}{{\cal F}}
\newcommand{\x}{{\cal O}}
\newcommand{\LL}{{\cal L}}

\def\mani{{\cal M}}
\def\calo{{\cal O}}
\def\calb{{\cal B}}
\def\calw{{\cal W}}
\def\calz{{\cal Z}}
\def\cald{{\cal D}}
\def\calc{{\cal C}}
\def\to{\rightarrow}
\def\ele{{\hbox{\sevenrm L}}}
\def\ere{{\hbox{\sevenrm R}}}
\def\zb{{\bar z}}
\def\wb{{\bar w}}
\def\nodiv{\mid{\hbox{\hskip-7.8pt/}}}
\def\menos{\hbox{\hskip-2.9pt}}
\def\dr{\dot R_}
\def\drr{\dot r_}
\def\ds{\dot s_}
\def\da{\dot A_}
\def\dga{\dot \gamma_}
\def\ga{\gamma_}
\def\dal{\dot\alpha_}
\def\al{\alpha_}
\def\cl{{closed}}
\def\cls{{closing}}
\def\vev{vacuum expectation value}
\def\tr{{\rm Tr}}
\def\to{\rightarrow}
\def\too{\longrightarrow}


\def\a{\alpha}
\def\b{\beta}
\def\c{\gamma}
\def\d{\delta}
\def\e{\epsilon}           
\def\f{\phi}               
\def\vf{\varphi}  \def\tvf{\tilde{\varphi}}
\def\vp{\varphi}
\def\g{\gamma}
\def\h{\eta}
\def\j{\psi}
\def\k{\kappa}                    
\def\l{\lambda}
\def\m{\mu}
\def\n{\nu}
\def\o{\omega}  \def\w{\omega}
\def\q{\theta}  \def\th{\theta}                  
\def\r{\rho}                                     
\def\s{\sigma}                                   
\def\t{\tau}
\def\u{\upsilon}
\def\x{\xi}
\def\z{\zeta}
\def\pt{\tilde{\varphi}}
\def\tt{\tilde{\theta}}
\def\lab{\label}
\def\6{\partial}
\def\wg{\wedge}
\def\atanh{{\rm arctanh}}
\def\bpsi{\bar{\psi}}
\def\bt{\bar{\theta}}
\def\bvf{\bar{\varphi}}

%

\newfont{\namefont}{cmr10}
\newfont{\addfont}{cmti7 scaled 1440}
\newfont{\boldmathfont}{cmbx10}
\newfont{\headfontb}{cmbx10 scaled 1728}
\newcommand{\re}{\,\mathbb{R}\mbox{e}\,}
\newcommand{\hyph}[1]{$#1$\nobreakdash-\hspace{0pt}}
\providecommand{\abs}[1]{\lvert#1\rvert}
\newcommand{\Nugual}[1]{$\mathcal{N}= #1 $}
\newcommand{\sub}[2]{#1_\text{#2}}
\newcommand{\partfrac}[2]{\frac{\partial #1}{\partial #2}}
\newcommand{\bsp}[1]{\begin{equation} \begin{split} #1 \end{split} \end{equation}}
\newcommand{\calF}{\mathcal{F}}
\newcommand{\calO}{\mathcal{O}}
\newcommand{\calM}{\mathcal{M}}
\newcommand{\calV}{\mathcal{V}}
\newcommand{\bbZ}{\mathbb{Z}}
\newcommand{\bbC}{\mathbb{C}}

\numberwithin{equation}{section}

\newcommand{\Tr}{\mbox{Tr}}    


%
\renewcommand{\theequation}{{\rm\thesection.\arabic{equation}}}
\begin{titlepage}
\vspace{0.1in}

\begin{center}
\Large \bf Fermionic impurities in Chern-Simons-matter theories
\end{center}
\vskip 0.2truein
\begin{center}
Paolo Benincasa  \footnote{paolo.benincasa@usc.es}
 and
Alfonso V. Ramallo\footnote{alfonso@fpaxp1.usc.es}\\
\vspace{0.2in}
\it{
Departamento de  F\'\i sica de Part\'\i  culas, Universidade
de Santiago de
Compostela\\and\\Instituto Galego de F\'\i sica de Altas
Enerx\'\i as (IGFAE)\\E-15782, Santiago de Compostela, Spain
}

\vspace{0.2in}
\end{center}
\vspace{0.2in}
\centerline{{\bf Abstract}}
We study the addition of quantum fermionic impurities to the ${\cal N}=6$ supersymmetric Chern-Simons-matter theories in $2+1$ spacetime dimensions. The impurities are introduced by means of Wilson loops in the antisymmetric representation of the gauge group. In a holographic setup, the system is represented by considering D6-branes probing the $AdS_4\times {\mathbb C}{\mathbb P}^3$ background of type IIA supergravity. We study the thermodynamic properties of the system and show how a Kondo lattice model with holographic dimers can be constructed. By computing the Kaluza-Klein fluctuation modes of the probe brane we determine the complete spectrum of dimensions of the impurity operators. A very rich structure is found, depending both on the Kaluza-Klein quantum numbers and on the filling fraction of the impurities.

\smallskip
\end{titlepage}
\setcounter{footnote}{0}

\tableofcontents

\section{Introduction}
The gauge/gravity correspondence \cite{jm,Aharony:1999ti} has been an invaluable analytic tool which has provided new insights on the the dynamics of gauge theories in the strong coupling regime. Indeed, in the framework of string theory, the gauge symmetry is realized as  the worldvolume symmetry  of a stack of coincident branes and the corresponding gravitational background encodes relevant properties of the gauge theory in the 't Hooft limit. The richness of string theory allows several possibilities to deform the theory by adding extra degrees of freedom. In particular, one can add new sets of branes which intersect with the color branes that host the gauge symmetry and create a defect in the gauge theory. By studying the dynamics of the extra branes in the original background one can extract useful dynamical information of the corresponding gauge theory with impurities.

In this paper we will concentrate on analyzing the case in which the defects are point-like. The canonical example of this type of systems is ${\cal N}=4$ Super-Yang-Mills (SYM) with fermionic impurities. The brane realization of this system is obtained by adding D5-branes to the $AdS_5\times S^5$ background, dual to ${\cal N}=4$  SYM. The D5-branes are extended in the holographic direction and wrap an ${\mathbb S}^4$ inside the ${\mathbb S}^5$ in such a way that their worldvolume is an $AdS_2\times {\mathbb S}^4$ submanifold of $AdS_5\times {\mathbb S}^5$ \cite{Pawelczyk:2000hy,Camino:2001at}.  The ${\mathbb S}^4$ that the D5-branes wrap are obtained by fixing the latitude inside the ${\mathbb S}^5$ to a discrete set of angles, which are obtained by imposing a quantization condition \cite{Camino:2001at}. These configurations can be regarded as describing bound states of fundamental strings extended along the holographic coordinate. Equivalently, they represent Wilson lines in the antisymmetric representation of the $SU(N)$ gauge group of the ${\cal N}=4$ SYM theory \cite{Yamaguchi:2006tq}. As shown in detail in \cite{Gomis:2006sb}, these Wilson loops can be obtained by integrating out a fermionic impurity, which is localized in the spatial gauge theory directions and whose occupation number is just the number of strings carried by the D5-brane. More recently, these configurations have been utilized \cite{Kachru:2009xf,Kachru:2010dk} to construct holographic dimer models, by considering D5-branes which connect two spatially separated impurities on the boundary of $AdS_5$. It has been also argued that this setup of D3- and D5-branes realizes holographically the maximally supersymmetric Kondo model \cite{Mueck:2010ja,Harrison:2011fs} (see also \cite{Faraggi:2011bb,Faraggi-dos}). 

The holographic study of gauge theories with impurities may be useful to understand the applications of the AdS/CFT correspondence to condensed matter theories. For this reason it is very important to have examples of such systems at our disposal, in which both sides of the correspondence are clearly identified and one can perform non-trivial tests of the duality. Following this line of thought,  we study  in this paper the addition of fermionic impurities to the ${\cal N}=6$ Chern-Simons-matter theory proposed by Aharony et al. (ABJM) in ref. \cite{Aharony:2008ug} (based on previous results by Bagger and Lambert \cite{BL} and Gustavsson \cite{Gustavsson:2007vu}). The ABJM theory describes the dynamics of multiple M2-branes at a ${\mathbb C}^4/{\mathbb Z}_k$ singularity and is a $U(N)\times U(N)$ (2+1)-dimensional Chern-Simons gauge theory with levels $(k,-k)$ and bifundamental matter fields. In the large $N$ limit this theory admits a supergravity description in M-theory in terms of the $AdS_4\times  {\mathbb S}^7/{\mathbb Z}_k$ geometry with flux. Moreover, 
when the Chern-Simons level $k$ is large  the system is better described in terms of type IIA supergravity. In this ten-dimensional description the geometry is of  the form $AdS_4\times {\mathbb C}{\mathbb P}^3$ with fluxes of the RR two- and four- forms. 

It was proposed in ref. \cite{Drukker:2008zx} that the Wilson lines in the antisymmetric representations of the gauge group of the ABJM theory are holographically realized as D6-branes extended along an $AdS_2\subset AdS_4$ and wrapping a five-dimensional submanifold of  ${\mathbb C}{\mathbb P}^3$. These are, therefore, the natural brane configurations to consider in order to construct the holographic dual of the Chern-Simons-matter theory with fermionic impurities. In this paper we will confirm this identification and we will study in detail some of the properties of the system with impurities. This brane setup constitutes the holographic description of a Kondo model in which the impurities are coupled to an ambient interacting CFT in three spacetime dimensions.

In the configurations  we will focus on, the probe D6-branes capture part of the RR flux of the background and, as a consequence, an electric worldvolume field must be turned on. This worldvolume field can be eliminated from the lagrangian of the probe by means of a Legendre transformation, after using the quantization condition found in \cite{Camino:2001at}. The result is an effective one-dimensional problem which can be solved by means of a first-order BPS differential equation which depends on the quantization integer $n$ (that represents the number of fundamental strings). In this paper we will mostly center in studying the simplest solutions of the BPS equation, which give rise to flux tubes made of $n$ strings with $0< n<N$  ($N$ is the rank of the $U(N)$ gauge groups). There are, however, more general solutions which are related to the baryon vertex of the ABJM theory. A first analysis of these baryonic solutions is relegated to an appendix. We will leave a more complete study for  future work.

In the flux tube solutions used to create the impurities, the embedding of the D6-branes is such that one of the angular coordinates that parametrize the  ${\mathbb C}{\mathbb P}^3$ is fixed to a constant which depends on the number $n$ of strings. In the dual field theory $n$ is just the number of defect fermions and the ratio $n/N$ is the filling fraction of the impurities.  The D6-branes wrap a squashed $T^{1,1}$ space inside the internal $ {\mathbb C}{\mathbb P}^3$. Recall that the $T^{1,1}$  space is a five-dimensional Sasaki-Einstein space whose metric can be represented as a $U(1)$ fibration of ${\mathbb S}^2\times {\mathbb S}^2$. In our case this space is internally deformed by squashing factors which preserve the fiber structure and depend non-trivially on the filling fraction of the impurities. 

Once the setup required to add impurities is clearly identified,  one can study many properties of the system. We will first analyze its thermodynamic properties by considering a probe brane in a black hole background dual to the ABJM theory at finite temperature. We will also study the configurations in which two flux tubes ending on two different points on the boundary  of $AdS_4$ are connected in the bulk and we will show that there exists a dimerization transition similar to the one described in refs. \cite{Kachru:2009xf,Kachru:2010dk}. In order to establish the stability of the brane configuration, we will analyze in detail the fluctuations of a brane probe at zero temperature. We will be able to decouple the different modes and read the conformal dimensions of the various dual operators. To carry out this analysis we will have to develop the harmonic calculus on the squashed $T^{1,1}$ space. As a result we will uncover a rich structure which depends both on the filling fraction of the impurities and on the quantum numbers of the Kaluza-Klein harmonics of the squashed $T^{1,1}$.  As happens in other conifold-like theories, the conformal dimensions will not be, in general, rational numbers \cite{Gubser:1998vd}. We will be able to characterize the normalizable modes which do correspond to impurity operators of rational dimensions. These operators should be the ones protected by the supersymmetry of the system.

The rest of this paper is organized as follows. In section \ref{ABJM} we briefly review the gravity dual of the ABJM theory in the type IIA supergravity. In section \ref{D6-SUSY} 
we study a class of configurations of the probe D6-branes in which they wrap a  co-dimension one submanifold of  $ {\mathbb C}{\mathbb P}^3$. Section \ref{flux-tubes} is devoted to analyzing the flux tube embeddings. The thermodynamic properties of these configurations are studied in section \ref{blackABJM} by considering embeddings in the black hole version of the ABJM metric. In this section we compute holographically the free energy and entropy of the impurity. In section \ref{dimers} we study D6-brane embeddings in which the brane is hanging from the boundary and ends on two different points. In this section we compute the free energy of this hanging tube configuration and we explore the corresponding dimerization transition. In sections \ref{Fluct} and \ref{dimer-fluct} we analyze in detail the fluctuation modes of the D6-branes around the flux tube and dimer configurations at zero temperature. We check their stability and compute the conformal dimensions of the dual impurity fields. In section \ref{conclusions} we summarize our results, extract some conclusions and point out some lines for possible future research. The paper is completed with several appendices. In appendix \ref{baryon} we briefly present the BPS worldvolume equation and find its  solutions. In appendix \ref{fluctuations} we derive the lagrangian for the fluctuations of the probe brane. In appendix \ref{harmonics} we study the laplacian for the squashed $T^{1,1}$. In appendix \ref{harmonic} we develop the harmonic calculus for these spaces. Finally, in appendix \ref{Heun} we review some basic facts about Heun's equation and its connection to a quantum integrable model.

\section{The ABJM background}
\label{ABJM}

The ABJM background \cite{Aharony:2008ug} of type IIA string theory is characterized by a ten-dimensional metric of the form $AdS_4\times {\mathbb C}\,{\mathbb P}^3$:
\beq
ds^2\,=\,{R^3\over 4k}\,(ds^2_{AdS_4}\,+\,4\,ds^2_{{\mathbb C}{\mathbb P}^3})\,\,,
\label{ads4-cp3-metric}
\eeq
where $ds^2_{AdS_4}$ and $ds^2_{{\mathbb C}{\mathbb P}^3}$ are, respectively, the 
$AdS_4$ and ${\mathbb C}{\mathbb P}^3$ metrics. The former can be written, in Poincar\'e coordinates, as:
\beq
ds^2_{AdS_4}\,=\,r^2\,\big(\,-dt^2+dx^2+dy^2\,\big)\,+\,{dr^2\over r^2}\,\,.
\label{AdS4-metric}
\eeq
In (\ref{ads4-cp3-metric}) we have represented the $AdS_4$ radius in terms of $R$ and $k$. Actually, the ABJM solution depends on two integers $N$ and $k$ which correspond, in the  dual gauge theory, to the rank of the gauge groups and to the Chern-Simons levels, respectively. The  $AdS_4$ radius in (\ref{ads4-cp3-metric}) can be written in terms of these two  integers as:
\beq
{R^3\over 4k}\,=\,\pi\,\sqrt{{2N\over k}}\,\,,
\label{Ads4-radius}
\eeq
where we are taking the Regge slope $\alpha'=1$.
The metric $ds^2_{{\mathbb C}{\mathbb P}^3}$ in (\ref{ads4-cp3-metric})
 is the  canonical Fubini-Study metric. In order to write its expression in a convenient form, let us consider four complex coordinates $z^i$ ($i=1,\cdots, 4$) such that $\sum_i\,|z^i|^2\,=\,1$. Therefore, these $z^i$'s parametrize a seven-sphere. We will represent them  as:
\bear
&&z^1\,=\,\cos{\alpha\over 2}\,\cos{\theta_1\over 2}\,
e^{{i\over 4}\,(2\varphi_1\,+\,\chi\,+\,\xi)}\,\,,\qquad
z^2\,=\,\cos{\alpha\over 2}\,\sin{\theta_1\over 2}\,
e^{{i\over 4}\,(-2\varphi_1\,+\,\chi\,+\,\xi)}\,\,,\rc\rc
&&z^3\,=\,\sin{\alpha\over 2}\,\sin{\theta_2\over 2}\,
e^{{i\over 4}\,(2\varphi_2\,-\,\chi\,+\,\xi)}\,\,,\qquad
z^4\,=\,\sin{\alpha\over 2}\,\cos{\theta_2\over 2}\,
e^{{i\over 4}\,(-2\varphi_2\,-\,\chi\,+\,\xi)}\,\,.
\eear
It is straightforward to verify that 
the metric of the seven sphere ${\mathbb S}^7$ can be written  in these coordinates as a $U(1)$ bundle over ${{\mathbb C}{\mathbb P}^3}$, namely:
\beq
ds^2_{{\mathbb S}^7}\,=\, \,ds^2_{{\mathbb C}{\mathbb P}^3}\,+\,(d\zeta+A)^2\,\,.
\eeq
In our parametrization $\zeta\sim\xi$ and the ${{\mathbb C}{\mathbb P}^3}$ metric $ds^2_{{\mathbb C}{\mathbb P}^3}$, as well as  the $U(1)$ connection $A$, are given by:
\bear
&&ds^2_{{\mathbb C}{\mathbb P}^3}\,=\,{1\over 4}\,\Big[\,d\alpha^2\,+\,
\cos^2{\alpha\over 2}\,\big(\,d\theta_1^2\,+\,\sin^2\theta_1\,d\varphi_1^2\,\big)\,+\,
\sin^2{\alpha\over 2}\,\big(\,d\theta_2^2\,+\,\sin^2\theta_2\,d\varphi_2^2\,\big)\,+\,
\rc\rc
&&\qquad\qquad\qquad\qquad
+\sin^2{\alpha\over 2}\,\cos^2{\alpha\over 2}\,
\big(\,d\chi+\cos\theta_1 d\varphi_1\,+\,\cos\theta_2 d\varphi_2\,\big)^2\,\,\Big]\,\,,\rc\rc
&&A\,=\,\cos\alpha\, d\chi\,+\,2\,\cos^2{\alpha\over 2}\,\cos\theta_1 d\varphi_1\,-\,
2\,\sin^2{\alpha\over 2}\,\cos\theta_2 \,d\varphi_2\,\,.
\label{CP3-folliated}
\eear
The ranges of the angles are $0\le \alpha,\theta_1, \theta_2, \le \pi$, 
$0\le \varphi_1, \varphi_2 <2\pi$ and $0\le \chi < 4\pi$.  Notice that, in the representation (\ref{CP3-folliated}), the submanifolds  with $\alpha={\rm constant}$  are squashed $T^{1,1}$ five-dimensional spaces, which foliate  the entire ${\mathbb C}{\mathbb P}^3$.

The ABJM type IIA background is also endowed with RR two- and four- forms $F_2$ and $F_4$, as well as a constant dilaton  $\phi$, which are given by:
\beq
e^{2\phi}\,=\,{R^3\over k^3}\,\,,\qquad
F_4\,=\,{3\over 8}\,R^3\,\,\Omega_{AdS_4}\,\,,\qquad
F_2\,=\,{k\over 4}\,dA\,\,,
\label{dilaton-forms-ABJM}
\eeq
where $\Omega_{AdS_4}$ is the volume element of the $AdS_4$ metric (\ref{AdS4-metric}). This background is a good supergravity dual of the Chern-Simons matter theory when the  $AdS_4$ radius is large in string units (and the corresponding curvature is small) and when the string coupling $e^{\phi}$ is small. From (\ref{Ads4-radius}) and (\ref{dilaton-forms-ABJM}) it follows \cite{Aharony:2008ug} that these conditions are satisfied  when  $N^{{1\over 5}}<<k<< N$.

In type IIA supergravity the RR six-form $F_6$ is defined as the Hodge dual of the RR four-form $F_4$. For the ABJM background  $F_6$ is given by:
\beq
F_6\,=\,{}^*\,F_4\,=\,{3R^6\over 2^8 k}\,\sin^3\alpha\,
d\alpha\wedge\epsilon_5\,\,,
\eeq
where $\epsilon_5$ is the five-form:
\beq
\epsilon_5\,=\,
\sin\theta_1\sin\theta_2\, d\theta_1\wedge d\theta_2\wedge d\chi\wedge d\varphi_1\wedge d\varphi_2\,\,.
\eeq
Let us now find the five-form potential $C_5$ for $ F_6$ (\ie\ such that  $F_6=dC_5$). 
This potential will be very useful in the analysis of the probe D6-branes of section \ref{D6-SUSY}.  First of all, we define the function $C(\alpha)$ as the solution of the following first-order equation:
\beq
{dC\over d\alpha}\,=\,-3\,\big(\sin\alpha\big)^3\,\,,\qquad\qquad
C(0)=0\,\,.
\eeq
This function is just:
\beq
C(\alpha)\,=\,\cos\alpha\big(\sin^2\alpha\,+\,2)\,-\,2\,\,.
\eeq
Then, it is immediate to verify that  the RR potential five-form $C_5$ can be taken as:
\beq
C_5\,=\,-{R^6\over 2^8\,k}\,C(\alpha)\,\epsilon_5\,\,.
\label{C5}
\eeq

\section{D6-brane probes}
\label{D6-SUSY}

Let us consider a D6-brane extended along the radial direction $r$ and wrapping a five-dimensional submanifold inside the ${\mathbb C}{\mathbb P}^3$ at fixed values of the Minkowski coordinates $x, y$. Notice that a configuration of this type creates a point-like defect in the gauge theory. In order to describe it in detail,  we will take the following worldvolume coordinates:
\beq
\zeta^{a}\,=\,(t,r,\gamma^i)\,=\,
(t,r,\theta_1, \varphi_1,\theta_2, \varphi_2, \chi)\,\,.
\label{wv-coordinates-flux}
\eeq
We will consider embeddings  in which the angle $\alpha$ is a function of the radial coordinate $r$, namely:
\beq
\alpha\,=\,\alpha(r)\,\,.
\eeq
Notice that the hypersurface $\alpha={\rm constant}$ defines a co-dimension one submanifold of the internal ${\mathbb C}{\mathbb P}^3$. In addition, we will switch on an electric worldvolume gauge field $F_{0r}$. Actually, such a worldvolume field is sourced by the RR potential $C_5$ through the Wess-Zumino (WZ) term of the D6-brane action (see below). The Dirac-Born-Infeld (DBI) piece of the brane action is just:
\beq
S_{DBI}\,=\,-T_{D_6}\,\int d^7 \zeta\, e^{-\phi}\,\sqrt{-\det (g+F)}\,\,,
\label{DBI-D6}
\eeq
where $g$ is the induced metric on the worldvolume of the D6-brane and $T_6$ is its tension.  The DBI determinant, integrated over the five angles $\gamma^i$  in (\ref{wv-coordinates-flux}) is:
\beq
T_{D_6}\,\int d^{5}\,\gamma\,e^{-\phi}\,\sqrt{-\det (g+F)}\,=\,{R^9\over 2^{10}\,\pi^3\,k^2}\,\,
\big(\sin\alpha \big)^3\,\,
\sqrt{1+r^2\,\big(\alpha'\big)^2\,-\,{16 k^2\over R^6}\,F_{0r}^2}\,\,,
\eeq
where $d^{5}\gamma=d\theta_1\wedge d\theta_2\wedge d\chi\wedge d\varphi_1\wedge d\varphi_2$  and $\alpha'$ denotes $d\alpha/dr$.  Let us define the effective DBI lagrangian density ${\cal L}_{DBI}$ as:
\beq
S_{DBI}\,=\,\int dt dr \,{\cal L}_{DBI}\,\,.
\eeq
Then, one has:
\beq
{\cal L}_{DBI}\,=\,-{R^9\over 2^{10}\,\pi^3\,k^2}\,\,
\big(\sin\alpha \big)^3\,\,
\sqrt{1+r^2\,\big(\alpha'\big)^2\,-\,{16 k^2\over R^6}\,F_{0r}^2}\,\,.
\label{L-DBI-SUSY}
\eeq
Moreover, the WZ action is:
\beq
S_{WZ}\,=\,T_6\,\int C_5\wedge F\,\equiv \int dr \,dt\,{\cal L}_{WZ}\,\,.
\eeq
By using the expression of $C_5$ written in (\ref{C5}), one arrives at the following expression of ${\cal L}_{WZ}$:
\beq
{\cal L}_{WZ}\,=\,-{R^6\over 2^8\,\pi^3\,k}\,C(\alpha)\,F_{0r}\,\,.
\label{L-WZ}
\eeq
Then, the total lagrangian density ${\cal L}\,=\,{\cal L}_{DBI}+{\cal L}_{WZ}$ can be written as:
\beq
{\cal L}\,=\,-{R^6\over 2^8\,\pi^3\,k}\,\Bigg[\,
{R^3\over 4k}\,\big(\sin\alpha\big)^3\,
\sqrt{1+r^2\,\big(\alpha'\big)^2\,-\,{16 k^2\over R^6}\,F_{0r}^2}
\,+\,
C(\alpha)\,F_{0r}\,\Bigg]\,\,.
\label{total-L-SUSY}
\eeq
The equation of motion of the gauge field derived from the lagrangian density (\ref{total-L-SUSY}) 
 implies that:
\beq
{\partial {\cal L}\over \partial F_{0r}}\,=\,{\rm constant}\,\,,
\eeq
which is nothing but the Gauss law. 
By using the quantization condition derived in \cite{Camino:2001at}, one can relate this constant to the number $n$ of strings (quarks) of the flux tube, namely:
\beq
{\partial {\cal L}\over \partial F_{0r}}\,=\,n\,T_f\,\,,
\label{quantization-condition}
\eeq
where  $T_f$ is the tension of the fundamental string  and  $n\in{\mathbb Z}$ is the fundamental string charge carried by the D6-brane. For our lagrangian density (\ref{total-L-SUSY}) we get:
\beq
{\partial {\cal L}\over \partial F_{0r}}\,=\,
{R^6\over 2^8\,\pi^3\,k}\,
\Bigg[\,{4k\over R^3}\,
{ F_{0r}\,\big(\sin\alpha\big)^3\over 
\sqrt{1+r^2\,\big(\alpha'\big)^2\,-\,{16 k^2\over R^6}\,F_{0r}^2}}\,-\,
C(\alpha)\,\Bigg]\,\,. 
\label{partialL-F_0r}
\eeq
By using this expression  in (\ref{quantization-condition}) we can determine $F_{0r}$ in terms of the other variables. Indeed, 
let us define the function ${\cal C}_n(\alpha)$ as:
\beq
{\cal C}_n(\alpha)\,\equiv C(\alpha)\,+\,{4n\over N}\,\,,
\eeq
whose explicit expression is:
\beq
{\cal C}_n(\alpha)\,=\,\cos\alpha\big(\sin^2\alpha\,+\,2)\,-\,2\,+\,
{4n\over N}\,\,.
\label{calCn}
\eeq
Notice that ${\cal C}_n(\alpha)$ satisfies the equation:
\beq
{d{\cal C}_n(\alpha)\over d\alpha}\,=\,-3\,
\big(\sin\alpha\big)^3\,\,,
\qquad\qquad
{\cal C}_n(0)\,=\,{4n\over N}\,\,.
\eeq
By combining (\ref{quantization-condition}) and (\ref{partialL-F_0r}) one can get a expression of $F_{0r}$ in terms of the function ${\cal C}_n(\alpha)$, namely:
\beq
F_{0r}\,=\,{R^3\over 4k}\,{
\sqrt{1\,+\,r^2\,\big(\alpha'\big)^2}\over
\sqrt{\big(\sin\alpha\big)^6\,+\,{\cal C}_n(\alpha)^2}}\,\,{\cal C}_n(\alpha)\,\,.
\label{F0r}
\eeq
In order to eliminate the electric field from the equations of motion, 
let us next compute the hamiltonian of the system by performing the Legendre transformation of the lagrangian density (\ref{total-L-SUSY}):
\beq
H\,=\,\int dr\,\Big[\,F_{0r}\,{\partial {\cal L}\over \partial F_{0r}}\,-\,{\cal L}\,
\Big]\,\equiv\,\int dr \,{\cal H}\,\,.
\eeq
By explicit calculation, one gets:
\beq
F_{0r}\,{\partial {\cal L}\over \partial F_{0r}}\,-\,{\cal L}\,=\,
{R^9\over 2^{10}\,\pi^3\,k^2}\,
\,\,
{1\,+\,r^2\,\big(\alpha'\big)^2
\over \sqrt{1\,+\,r^2\,\big(\alpha'\big)^2\,-\,{16 k^2\over R^6}\,F_{0r}^2
}}
\,\,\,\big(\sin\alpha\big)^3\,\,.
\eeq
Moreover, from the expression of $F_{0r}$ written in (\ref{F0r}), we deduce:
\beq
{\big(\sin\alpha\big)^3\over 
\sqrt{1\,+\,r^2\,\big(\alpha'\big)^2\,-\,{16 k^2\over R^6}\,F_{0r}^2}
}\,=\,
{\sqrt{\big(\sin\alpha\big)^6\,+\,{\cal C}_n(\alpha)^2}
\over 
\sqrt{1\,+\,r^2\,\big(\alpha'\big)^2}}\,\,.
\eeq
Thus, the hamiltonian density ${\cal H}$ can be written as:
\beq
{\cal H}\,=\,
{R^9\over 2^{10}\,\pi^3\,k^2}\,
\,\,
\sqrt{1\,+\,r^2\,\big(\alpha'\big)^2}\,\,
\sqrt{\big(\sin\alpha\big)^6\,+\,{\cal C}_n(\alpha)^2}\,\,.
\label{hamiltonian_density}
\eeq
The different embeddings of the D6-brane described by our ansatz can be found by integrating the Euler-Lagrange equation derived from (\ref{hamiltonian_density}). In the main part of this paper we will concentrate on studying the simplest of these configurations, namely those for which $\alpha$ is a constant. These embeddings will be  analyzed in detail in following sections. In appendix \ref{baryon} we will present embeddings with non-constant $\alpha$, which satisfy a first-order BPS equation.  These last configurations are related to the baryon vertex of the ABJM theory (see \cite{Callan:1998iq,Callan:1999zf} for a similar analysis in the $AdS_5\times S^5$ background). We will leave their detailed study for a future work.

\section{Flux tube configurations}
\label{flux-tubes}

Let us now look for solutions of the equations of motion which have a constant $\alpha$ angle. Clearly, we should require that:
\beq
{\partial {\cal H}\over \partial \alpha}\Big|_{\alpha'=0}\,=\,0\,\,.
\eeq
Therefore, we need to compute the derivative of the second square root in (\ref{hamiltonian_density}). With this purpose in mind, we define the following function:
\beq
\Lambda_n(\alpha)\,\equiv\, \big(\sin\alpha\big)^2\,\cos\alpha\,-\,
{\cal C}_n(\alpha)\,\,.
\label{Lambda_n-def}
\eeq
As:
\beq
{d\over d\alpha}\,\,\Big[\,
\big(\sin\alpha\big)^6\,+\,{\cal C}_n(\alpha)^2\,\Big]\,=\,6\,
\big(\sin\alpha\big)^3\,\Lambda_n(\alpha)\,\,,
\eeq
we get:
\beq
{\partial {\cal H}\over \partial \alpha}\Big|_{\alpha'=0}\,=\,
{3\,R^9\over 2^{10}\,\pi^3\,k^2}
\,\,
{\big(\sin\alpha\big)^3\over 
\sqrt{\big(\sin\alpha\big)^6\,+\,{\cal C}_n(\alpha)^2}}\,\,
\Lambda_n(\alpha)\,\,.
\eeq
Therefore, the non-trivial configurations with constant $\alpha$ occur when $\alpha=\alpha_n$, where $\alpha_n$ is a solution of the  equation:
\beq
\Lambda_n(\alpha_n)\,=\,0\,\,.
\label{minimum_condition}
\eeq
Let us now study the solutions of this equation. By plugging the definition of ${\cal C}_n(\alpha)$  written in (\ref{calCn}) into  (\ref{Lambda_n-def}),  one can obtain  the explicit expression of $\Lambda_n(\alpha)$, which  is given by:
\beq
\Lambda_n(\alpha)\,=\,
2\,\,\Big(\,1\,-\,{2n\over N}\,-\,\cos\alpha\,\Big)\,\,.
\label{Lambda_n}
\eeq
Thus, the roots of (\ref{minimum_condition}) are:
\beq
\cos\alpha_n\,=\,1\,-\,{2n\over N}\,\,.
\label{cos_alpha}
\eeq
Notice that there is a unique solution $\alpha_n\in [0,\pi]$ for every $n$ in the range $0\le n\le N$. This configuration was proposed in reference \cite{Drukker:2008zx} as the holographic dual of the Wilson loop of the ABJM theory in the antisymmetric representations of the gauge group. The fact  that $n\le N$ supports this identification and is a manifestation of the so-called stringy exclusion principle.  The energy density for these configurations is just:
\beq
E_n\,\equiv\,
{\cal H}(\alpha=\alpha_n)\,=\,{N\over 8\pi}\,{R^3\over 4k}\,\,
\big(\sin\alpha_n\big)^2\,\,.
\eeq
Using the value of  $\alpha_n$ written in (\ref{cos_alpha}), we get the following binomial law for the energy density:
\beq
E_n\,=\,\sqrt{\lambda \over 2}\,\,{n(N-n)\over N}\,\,,
\label{E_n}
\eeq
where we have written the result in terms of the 't Hooft coupling  $\lambda$ of the Chern-Simons-matter theory ($\lambda=N/k$). Notice that $E_n$ is invariant under the change $n\to N-n$, as it should for the holographic dual of an object transforming in the anti-symmetric representation of the gauge group. Notice also that for small $n$ (or large $N$) the energy density $E_n$ is just equal to $n$ times the tension of a fundamental string extended along the radial direction in the  geometry (\ref{ads4-cp3-metric}) (which is given by $T_f\,R^3/4k\,=\,\sqrt{\lambda/2}$). This is, again, in favor of the interpretation of this configuration as a bound state of fundamental strings, as those studied in \cite{Camino:2001at} for backgrounds generated by Dp-branes. The fact that $E_n\le n T_f\,R^3/4k$ is indicating  that the formation of the bound state is energetically favored and that the configuration is stable. We will check explicitly this stability by computing the fluctuation spectra of the brane in section \ref{Fluct}. It is also interesting to point out here that, as checked explicitly in \cite{Drukker:2008zx}, these configurations are 1/6 BPS, \ie\ they preserve four real supercharges.

The constant electric field $\bar f_{0r}$ corresponding to this configuration  can be obtained from (\ref{F0r}) by substituting $\alpha$ by $\alpha_n$, namely:
\beq
\bar f_{0r}\,=\,{R^3\over 4k}\,\,\cos\alpha_n\,\,.
\label{f-unperturbed}
\eeq
Notice that the induced metric on the D6-brane worldvolume takes the form:
\beq
d s^2\,=\,
{R^3\over 4k}\,\,\Big[\,-\,r^2\,dt^2\,+\,{dr^2\over r^2}\,+\,d s^2_{\tilde T^{1,1}}\,\Big]\,\,,
\label{induced-metric-unperturbed}
\eeq
which is of the form $AdS_2\times \tilde T^{1,1}$, 
with $\tilde T^{1,1}$ being a squashed $T^{1,1}$ space, with metric:
\bear
&&d s^2_{\tilde T^{1,1}}\,=\,\tilde g_{ij}\,d\gamma^i\,d\gamma^j\,=\,
\cos^2{\alpha_n\over 2}\,\big(\,d\theta_1^2\,+\,\sin^2\theta_1\,d\varphi_1^2\,\big)\,+\,
\sin^2{\alpha_n\over 2}\,\big(\,d\theta_2^2\,+\,\sin^2\theta_2\,d\varphi_2^2\,\big)\,+\,
\rc\rc
&&\qquad\qquad\qquad\qquad\qquad\qquad
+\sin^2{\alpha_n\over 2}\,\cos^2{\alpha_n\over 2}\,
\big(\,d\chi+\cos\theta_1 d\varphi_1\,+\,\cos\theta_2 d\varphi_2\,\big)^2\,\,\,.
\label{squashedT11}
\eear
Taking into account  that the angles $\alpha_n$ in (\ref{cos_alpha}) satisfy:
\beq
\sin^2{\alpha_n\over 2}\,=\,{n\over N}\,\,,
\qquad\qquad
\cos^2{\alpha_n\over 2}\,=\,{N-n\over N}\,\,,
\eeq
it follows that  the metric  $d s^2_{\tilde T^{1,1}}$ can be written in  the form:
\bear
&&d s^2_{\tilde T^{1,1}}\,=\,
{N-n\over N}\,\,\big(\,d\theta_1^2\,+\,\sin^2\theta_1\,d\varphi_1^2\,\big)\,+\,
{n\over N}\,\,\big(\,d\theta_2^2\,+\,\sin^2\theta_2\,d\varphi_2^2\,\big)\,+\,
\rc\rc
&&\qquad\qquad\qquad\qquad\qquad\qquad
+\,{n(N-n)\over N^2}\,
\big(\,d\chi+\cos\theta_1 d\varphi_1\,+\,\cos\theta_2\, d\varphi_2\,\big)^2\,\,\,.
\label{squashedT11-n}
\eear
Notice that the line element in (\ref{squashedT11-n}) collapses to the one of to a two-sphere when $n=0,N$. This implies that the corresponding D6-brane configurations are singular and, accordingly, we will assume from now on that $0<n<N$.  Moreover, the metric (\ref{squashedT11-n}) displays the following symmetry:
\beq
(\theta_1,\varphi_1)\,\,\leftrightarrow \,\,(\theta_2,\varphi_2)\,\,,
\qquad\qquad\qquad
n\,\,\leftrightarrow \,\,N-n\,\,.
\label{n-N-symmetry}
\eeq
The configuration of D6-branes described above creates a point-like defect in the bulk Chern-Simons matter theory. The $n$ fundamental strings  carried out by the D6-branes introduce  $n$ fermions localized on the defect, which can be regarded as fermionic impurities in the Chern-Simons-matter theory. As pointed out in \cite{Gomis:2006sb} for the similar configurations in ${\cal N}=4$ SYM, by integrating out the defect fermions one should get an effective theory for the bulk fields in which there is an insertion of a Wilson loop in the antisymmetric representation of the gauge group. This argument establishes a correspondence between the Wilson loop interpretation of these configurations (as proposed in \cite{Drukker:2008zx}) and the model of quantum impurities developed in this paper.  Notice that our setup describes a Kondo model in which the ambient theory is an interacting supersymmetric  Chern-Simons-matter theory. The ratio:
\beq
\nu\,=\,{n\over N}\,\,,
\eeq
will be referred to as the filling fraction. Notice that the metric (\ref{squashedT11-n}) depends on $\nu$ and that under the symmetry (\ref{n-N-symmetry}) the filling fraction transforms as $\nu\to 1-\nu$ and, therefore, (\ref{n-N-symmetry}) can be regarded as the realization, in our holographic model,  of the particle-hole transformation. It is also clear from (\ref{n-N-symmetry}) that this particle-hole transformation is realized geometrically by the exchange of the two ${\mathbb S}^2$'s inside the $\tilde T^{1,1}$. 
The properties of the defect operators depend on $\nu$ and, in our setup,  they will be encoded in the squashing of the internal $\tilde T^{1,1}$ geometry. 

In what follows we will study holographically the properties of the impurity model. We will start in the next section by analyzing its basic thermodynamical properties.

\section{Impurities at finite temperature}
\label{blackABJM}

Let us now consider the ABJM theory at finite temperature. The corresponding background is obtained from the supersymmetric one by substituting the $AdS_4$ factor by  the geometry of a black hole in   $AdS_4$, namely:
\beq
ds^2\,=\,{R^3\over 4k}\,(ds^2_{BH_4}\,+\,4\,ds^2_{{\mathbb C}{\mathbb P}^3})\,\,,
\label{BHmetric}
\eeq
with:
\beq
ds^2_{BH_4}\,=\,-r^2\,f(r)\,dt^2\,+\,{dr^2\over r^2 f(r)}\,+\,
r^2\,\big[\,dx^2+dy^2\,\big]\,\,,
\eeq
where $f(r)$ is the function:
\beq
f(r)\,=\,1\,-\,{r_0^3\over r^3}\,\,,
\label{blackening}
\eeq
and $r_0$ is related to the temperature of the black hole by means of the expression:
\beq
T\,=\,{1\over 2\pi}\,\,
\Big[\,{1\over \sqrt{g_{rr}}}\,\,{d\over dr}\,
\Big(\,\sqrt{\,-g_{tt}}\,\Big)\,
\Big]_{r=r \,_0}\,=\,
{3r_0\over 4\pi}\,\,.
\label{T-r0}
\eeq

In order to compute the free energy and entropy of the flux tube configuration described in section \ref{flux-tubes},  let us evaluate the Euclidean action of one of such configurations that extends from the horizon at $r=r_0$ until a cutoff value of the radial coordinate $r=r_{cutoff}$. The result is:
\beq
I_{D6}\,=\, {E_{n}\over T}\,\,\big(\,r_{cutoff}-r_0\,\big)\,\,,
\label{D6-euclidean_action_bare}
\eeq
where  $E_{n}$ is the tension written in (\ref{E_n}) and we have integrated over a periodic Euclidean time circle of period $1/T$.   The action (\ref{D6-euclidean_action_bare})  is divergent  as $r_{cutoff}\to\infty$ and must be renormalized by subtracting a counterterm. Following the holographic renormalization approach \cite{Skenderis:2002wp,Karch:2005ms}, the counterterm is just the action of the brane embedded in the metric (\ref{BHmetric}) at zero temperature  and extended from the origin at $r=0$ until $r=r_{cutoff}$, namely:
\beq
I_{D6}^{counter}\,=\,{ E_{n}\over T\,'}\,r_{cutoff}\,\,,
\label{D6-euclidean_action_counter}
\eeq
where $T\,'$ is given by:
\beq
{1\over T\,'}\,=\,{f(r_{cutoff})\over T}\,\,,
\eeq
and $f(r)$ is the function defined in (\ref{blackening}). To evaluate (\ref{D6-euclidean_action_counter}) we have chosen the time period $1/T\,'$ in such a way that it corresponds to the same induced metric at $r=r_{cutoff}$ as the one of the brane embedded in the black hole metric (\ref{BHmetric}) \cite{Kachru:2009xf}. Therefore, 
the renormalized euclidean action is just:
\beq
I_{D6}^{renormalized}\,=\,\lim_{r_{cutoff}\to \infty}\,\,
\Big(\,I_{D6}\,-\,I_{D6}^{counter}\,\Big)\,=\,
-\,{E_{n}\over T}\,\,r_0\,\,.
\eeq
In terms of this renormalized action,  the impurity free energy is  defined as:
\beq
F^{imp}\,=\,T\,I_{D6}^{renormalized}\,\,.
\eeq
Therefore, by using the value of the renormalized action, we get:
\beq
F^{imp}\,=\,-E_{n}\,\,r_0\,\,.
\label{Fimp-En}
\eeq
If one ignores non-abelian interactions, the free energy for $M$ D6-branes is just $M$ times the result (\ref{Fimp-En}). By using (\ref{E_n}) and (\ref{T-r0}), one gets in this case:
\beq
F^{imp}\,=\,-{2\pi\sqrt{2}\over 3}\,\,\sqrt{\lambda}\,\,n\,M\,
\Big(\,1\,-\,{n\over N}\,\Big)\,T\,\,.
\label{Fimp-n-T}
\eeq 
The impurity entropy is related to the free energy as:
\beq
{\cal S}^{imp}\,=\,-{\partial 
F^{imp}\over \partial T}\,\,.
\eeq
In our case, we get:
\beq
{\cal S}^{imp}\,=\,
{2\pi\sqrt{2}\over 3}\,\,\sqrt{\lambda}\,\,n\,M\,
\big(\,1\,-\,\nu\,\big)\,\,.
\label{imp-entropy}
\eeq
Notice that the impurity entropy (\ref{imp-entropy}) depends on (square root) of  the 't Hooft coupling $\lambda$, as also happens for the  ${\cal N}=4$ impurities \cite{Kachru:2009xf,Harrison:2011fs}. This fact is a consequence of the interacting nature of the ambient CFT. As a function of the filling fraction $\nu=n/N$, the entropy (\ref{imp-entropy}) is maximized at the half-filling point $\nu=1/2$, which is the fixed point of the particle-hole symmetry. This behavior is similar to the one found in the ${\cal N}=4$ theory \cite{Kachru:2009xf,Harrison:2011fs}. Notice, however, that the dependence on $\nu$ in our case is analytic.

\section{Hanging flux tubes}
\label{dimers}

In this section we will consider configurations in which the flux tube does not reach the origin of the holographic coordinate but instead it starts from the boundary of $AdS_4$, reaches a minimal value of $r$ and comes back to the boundary. We will see that, in order to have such a hanging string configuration, one must have a  non-constant cartesian coordinate  for the embedding.  Therefore, these connected configurations give rise  to dimers on the boundary field theory, \ie\  to flux tubes connecting $n$ quarks and 
$n$ antiquarks separated a finite distance.

Let us therefore consider a D6-brane  probe  as in section \ref{D6-SUSY} and let us choose the same set of worldvolume coordinates as in (\ref{wv-coordinates-flux}). We 
embed the D6-branes in the black hole geometry (\ref{BHmetric}) and we consider an ansatz in which both $\alpha$ and the cartesian coordinate $x$ vary with $r$, namely:
\beq
\alpha\,=\,\alpha(r)\,\,,\qquad\qquad
x\,=\,x(r)\,\,.
\label{dimer-ansatz}
\eeq
The DBI lagrangian density is now given by:
\beq
{\cal L}_{DBI}\,=\,-{R^9\over 2^{10}\,\pi^3\,k^2}\,\,
\big(\sin\alpha \big)^3\,\,
\sqrt{1+r^2\,f(r)\,\big(\alpha'\big)^2\,+\,r^4\,f(r)\,\big(x'\big)^2\,
-\,{16 k^2\over R^6}\,F_{0r}^2}\,\,.
\eeq
The WZ lagrangian is still given by (\ref{L-WZ}). Thus, the total lagrangian density is:
\beq
{\cal L}\,=\,-{R^6\over 2^8\,\pi^3\,k}\,\Bigg[\,
{R^3\over 4k}\,\big(\sin\alpha\big)^3\,
\sqrt{1+\,r^2\,f(r)\,\big(\alpha'\big)^2\,+\,r^4\,f(r)\,\big(x'\big)^2
\,-\,{16 k^2\over L^4}\,F_{0r}^2}
\,+\,
C(\alpha)\,F_{0r}\Bigg]\,\,.\qquad\qquad
\eeq
We can now eliminate $F_{0r}$ following the same steps as in section \ref{D6-SUSY} 
(\ie\ by integrating the Gauss law and by performing a Legendre transformation). Clearly, the result that is obtained in this way is the same as in section \ref{D6-SUSY}, after performing the substitution:
\beq
r^2\big(\alpha'\big)^2\,\to\,r^2\,f(r)\,\big(\alpha'\big)^2\,+\,r^4\,f(r)\,\big(x'\big)^2\,\,.
\eeq
Therefore, the worldvolume  electric gauge field is given by:
\beq
F_{0r}\,=\,{R^3\over 4k}\,{
\sqrt{1\,+\,r^2\,f(r)\,\big(\alpha'\big)^2\,+\,r^4\,f(r)\,\big(x'\big)^2}\over
\sqrt{\big(\sin\alpha\big)^6\,+\,{\cal C}_n(\alpha)^2}}\,\,{\cal C}_n(\alpha)
\label{F0r-haging}\,\,,
\eeq
where ${\cal C}_n(\alpha)$ is the same function as in (\ref{calCn}).  After the Legendre transform we now arrive at the following routhian:
\beq
{\cal H}\,=\,
{R^9\over 2^{10}\,\pi^3\,k^2}\,
\,\,
\sqrt{1\,+\,r^2\,f(r)\,\big(\alpha'\big)^2\,+\,r^4\,f(r)\,\big(x'\big)^2}\,\,
\sqrt{\big(\sin\alpha\big)^6\,+\,{\cal C}_n(\alpha)^2}\,\,.
\label{routhian_density}
\eeq
It follows that in this case there still exist solutions with constant $\alpha$ for the same angles $\alpha=\alpha_n$  as in (\ref{cos_alpha}). From now on we will restrict ourselves to the case in which $\alpha=\alpha_n$.  In this case:
\beq
{\cal H}\,=\,E_n\,\sqrt{1+r^4\,f(r)\,(x')^2}\,\,,
\label{calH-fixed-alpha}
\eeq
where $E_n$ is just the energy written in (\ref{E_n})  for the $\alpha=\alpha_n$  configuration with $x$ constant.  The equation of motion of $x(r)$ has a first integral given by:
\beq
{\partial {\cal H}\over \partial x'}\,=\,\Lambda\,\,,
\label{first-integral}
\eeq
where $\Lambda$ is a constant of integration. Then, one has:
\beq
E_n\,\,{r^4\,f(r)\,x'\over \sqrt{1+r^4\,f(r)\,(x')^2}}\,=\,\Lambda\,\,.
\label{xprime-implicit}
\eeq
Clearly, having $\Lambda=0$ in (\ref{xprime-implicit})  means that $x(r)$ is not constant. From this expression we can obtain $x'$ as a function of $r$:
\beq
x'\,=\,\pm\,{\hat \Lambda\over r^2\,\sqrt{f(r)}}\,\,
{1\over \sqrt{r^4\,f(r)\,-\,\hat\Lambda^2}}\,\,,
\label{xprime}
\eeq
where $\hat\Lambda$ is defined as:
\beq
\hat\Lambda\,\equiv\,{\Lambda\over E_n}\,\,.
\eeq
Notice that, as:
\beq
\sqrt{1+r^4\,f(r)\,\big(x'\big)^2}\,=\,
{r^2\sqrt{f(r)}\over \sqrt{r^4\,f(r)\,-\,\hat \Lambda^2}}\,\,,
\label{sqrt-xprime}
\eeq
the  electric worldvolume field $F_{0r}$ is given by the following function of the radial coordinate:
\beq
F_{0r}\,=\,{R^3\over 4k}\,\,{r^2\sqrt{f(r)}\over \sqrt{r^4\,f(r)\,-\,\hat \Lambda^2}}\,
\cos\alpha_n\,\,.
\label{electric-field-dimer}
\eeq

It is important to point out  that the two signs in (\ref{xprime}) are needed in order to describe a  hanging D6-brane that comes from $r=\infty$ until a minimal $r$ and then comes back to the boundary. The actual minimal value of $r$, \ie\ the radial coordinate of the turning point, is the solution $r_*$ of the following quadratic equation:
\beq
r_*^4\,-\,r_0^3\,r_*\,=\,\hat\Lambda^2\,\,.
\label{turning}
\eeq
By integrating the equation (\ref{xprime}) for $x(r)$, we get:
\beq
x(r)\,=\,\pm\,\hat\Lambda\,\int_{r_*}^{r}\,
{d\rho\over \sqrt{\rho(\rho^3-r_0^3)}\,
\sqrt{\rho(\rho^3-r_0^3)\,-\,\hat \Lambda^2}}\,\,.
\eeq
Let us introduce in this integral a new variable $z$, defined as $z=\rho/r_0$. One gets: \beq
x(r)\,=\,\pm\,{q\over r_0}\,\,
\int_{z_*}^{{r\over r_0}}\,{dz\over 
\sqrt{z^4-z}\,\sqrt{z^4-z-q^2}}\,\,,
\label{x-r-z}
\eeq
where $q$ is defined as:
\beq
\label{qABJM}
q\,\equiv\, {\hat \Lambda\over r_0^2}\,\,,
\eeq
and $z_*$ is the root of the following quartic equation:
\beq
z_*^4\,-\,z_*\,-q^2\,=\,0\,\,. 
\label{z*-q}
\eeq
Notice that $z_*=r_*/r_0$  and that $q$ parametrizes, through the relation (\ref{z*-q}), the radial position of the turning point.  Moreover, when $q=0$ eq. (\ref{z*-q}) can be solved to give 
$z_*=1$ (or $r_*=r_0$). Therefore, it follows from (\ref{qABJM}) and (\ref{xprime}) that $x$ is constant for $q=0$, which implies that these configurations correspond to branes that go straight from the boundary to the horizon of the black hole. Coming back to the general connected configuration,  let us suppose that the brane starts at $x=-L/2$ at $r=\infty$ and comes back to the boundary at $x=+L/2$ (\ie\ $L$ is the separation of the $n$-quark and the $n$-antiquark). Then from (\ref{x-r-z})  one gets that $L$ as a function of $q$ is given by :
\beq
L\,=\,{2q\over r_0}\,
\int_{z_*}^{\infty}\,{dz\over \sqrt{(z^4-z)\,(z^4-z-q^2)}}\,\,.
\label{L-q}
\eeq
\begin{figure}[htbp]
 \centering%
 {\scalebox{.55}{\includegraphics{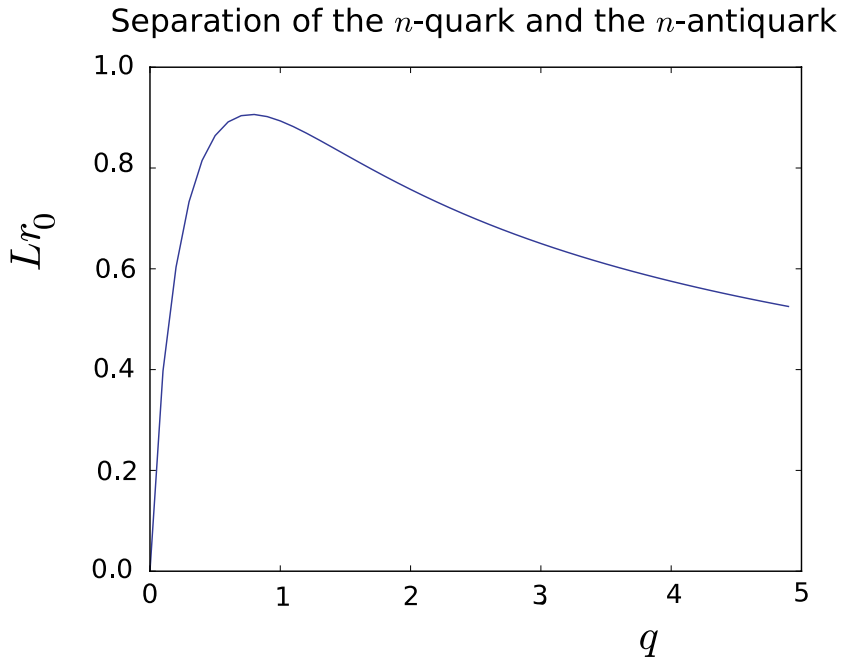}}}
 \caption{In this figure we plot $L\,r_0$ as a function of $q$, as given by (\ref{L-q}). The maximal value of $L\,r_0$ occurs when $q\approx 0.779$. For this value of $q$ we get $L\, r_0\approx 0.906$. }
 \label{Separation}
\end{figure}
In figure \ref{Separation} we have plotted $r_0\,L\,=\,{4\pi\over 3}\,\,T\,L$ as a function of $q$. It is clear from this figure that there exists a maximal value of $L$ for every temperature $T$ and that for fixed $L$ there are no connected configurations for large enough temperature. Moreover, when $r_0\,L$ is small enough  there are two possible values of $q$ which, according to (\ref{z*-q}), correspond to two possible values of $r_*$. In order to determine which of these two configurations is the dominant one, let us compute the free energy as a function of $q$. First of all, notice that plugging eq. (\ref{sqrt-xprime}) into the expression of the energy density given by (\ref{calH-fixed-alpha}), we get:
\beq
{\cal H}\,=\,E_n\,
{r^2\sqrt{f(r)}\over \sqrt{r^4\,f(r)\,-\,\hat \Lambda^2}}\,\,.
\eeq
Let us integrate this energy density for a  connected configuration whose ends are located at $r=r_{\max}$. One gets:
\beq
E\,=\,2 E_n\,\int_{r_*}^{r_{max}}\,
{\rho^2\sqrt{f(\rho)}\over \sqrt{\rho^4\,f(\rho)\,-\,\hat \Lambda^2}}\,d\rho\,\,.
\label{E-rho-dimer}
\eeq
Changing variables  in (\ref{E-rho-dimer}) from $\rho$ to $z=\rho/r_0$,  we get:
\beq
E\,=\,2 E_n\,r_0\,\int_{z_*}^{z_{max}}\,
{\sqrt{z^4-z}\over \sqrt{z^4-z-q^2}}\,\,
dz\,\,.
\eeq
This integral is divergent  when $z_{max}\to \infty$. In order to regulate it and get the free energy, we follow the procedure in \cite{Maldacena:1998im,Rey:1998ik} and subtract the energy corresponding to a disconnected configuration that reaches the point $r=r_{max}$. This energy is:
\beq
2 E_n\int_0^{r_{max}}\, d\rho\,=\,2 E_n\,r_0\Big[\,
\int_{z_*}^{z_{max}}\,dz\,+\,z_*\,\Big]\,\,.
\eeq
The regulated free energy for the connected configuration (after sending $z_{\max}\to\infty$) is given by:
\beq
F\,=\,2E_n\,r_0\,\Bigg[\,\int_{z_*}^{\infty}\,
\Bigg(\,{\sqrt{z^4-z}\over \sqrt{z^4-z-q^2}}\,-\,1\,\Bigg)\,dz\,-\,z_*\,\Bigg]\,\,.
\label{F-hanging}
\eeq
Notice that for $q=0$ (and $z_*=1$) the free energy in (\ref{F-hanging}) $F=-2E_n\,r_0$, \ie\ twice the result in (\ref{Fimp-En}) and (\ref{Fimp-n-T}), which confirms that this  $q=0$ configuration corresponds to two straight disconnected branes stretching from the boundary to the horizon. 
\begin{figure}[htbp]
 \centering%
 {\scalebox{.50}{\includegraphics{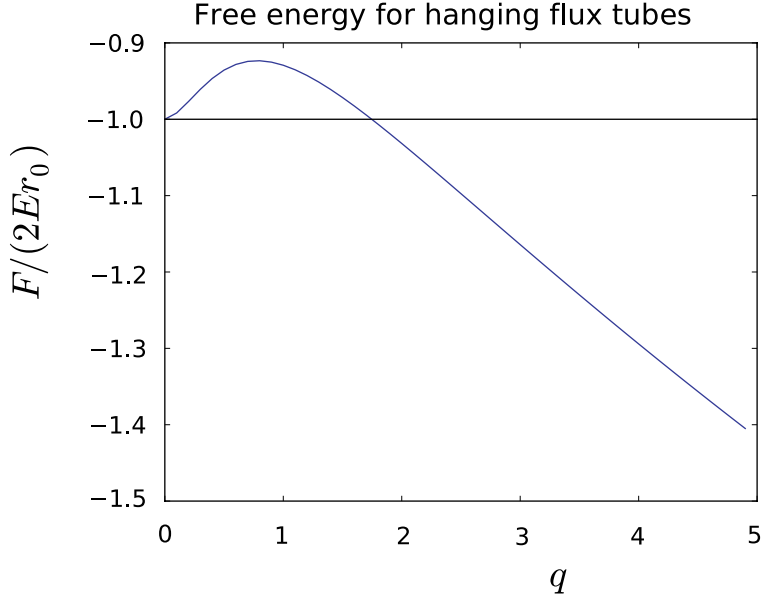}}}
 \caption{Plot of $F/(2E_n r_0)$ as a function of the parameter $q$. The curve reaches a maximum at $q\approx 0.789$, where it takes the value $F/(2E_n r_0)\approx  -0.923$. The dimerization transition starts at $q\approx 1.747$, where $F=-2E_n \,r_0$.}
 \label{feABJM}
\end{figure}
The free energy $F$ as a function of $q$ has been plotted in figure \ref{feABJM}.   By comparing figures \ref{Separation} and \ref{feABJM} it is easy to  verify that, when $r_0\,L$ is small enough to allow two solutions for $q$, the configuration with higher $q$ (\ie\ the one that penetrates less deeply in the bulk) is the one that has lower free energy and, therefore, is the dominant one.  It  is also clear from figure \ref{feABJM} that, for low enough temperature, this configuration has a free energy lower than  $-2E_n\,r_0$ and, therefore, it is also dominant over the disconnected one.  This reconnection of branes at low temperature is the dimerization transition described in \cite{Kachru:2009xf}. Its characteristics are very similar to the ones of the ${\cal N}=4$ four-dimensional theory and will not be elaborated further here. 

\section{Impurity fluctuations}
\label{Fluct}

Let us now consider fluctuations around the static configurations studied above. We will allow to fluctuate the coordinates transverse to the D6-brane (namely  the angle $\alpha$ and the cartesian coordinates $x$ and $y$), as well as the worldvolume field $F$. Thus, we write:
\beq
\alpha\,=\,\alpha_n\,+\,\xi\,\,,\qquad
F\,=\,\bar f\,+\,f\,\,,\qquad
x\,=\,\bar x\,+\,\chi\,\,,
\label{perturbation}
\eeq
where $\alpha_n$ is the angle determined in (\ref{cos_alpha}), $\bar x$ is the constant cartesian coordinate of the unperturbed D6-brane and the only non-zero component of the gauge field $\bar f$ is the one written in (\ref{f-unperturbed}). The equations of motion  for the fluctuations can be written in terms of the 
the open string metric (see appendix \ref{fluctuations}), which  takes the form:
\beq
{\cal G}_{mn}\,d\zeta^{m}\,d\zeta^{n}\,=\,{R^3\over 4k}\,\,
\big[\,\sin^2\alpha_n\,ds^2_{AdS_2}\,+\,d s^2_{\tilde T^{1,1}}\,\big]\,\,,
\eeq
where $ds^2_{AdS_2}=-r^2\,dt^2\,+\,dr^2/r^2$ and $d s^2_{\tilde T^{1,1}}$ has been written in (\ref{squashedT11}) and (\ref{squashedT11-n}).  In terms of the metric ${\cal G}_{mn}$, the total lagrangian density for the fluctuations  takes the form:
\bear
&&{\cal L}\,=\,-T_{6}\,\,{R^9\over 2^7\,k^2}\,\sin\alpha_n\,\sqrt{\tilde g}\,
\Bigg[\,{R^3\over 4k}\,{1\over 2}\,r^2\,{\cal G}^{mn}\,\partial_m\,\chi\,\partial_n\,\chi\,+\,
{R^3\over 4k}\,{1\over 2}\,{\cal G}^{mn}\,\partial_m\,\xi\,\partial_n\,\xi\,+\,\rc\rc
&&\qquad\qquad\qquad\qquad
+{1\over 4}\,{\cal G}^{mp}\,{\cal G}^{nq}\,f_{pq}\,f_{mn}\,-\,
{3\over 2}\,{\xi^2\over \sin^2\alpha_n}\,-\,{12k\over R^3}\,
{\xi f_{0r}\over  \sin^3\alpha_n}\,\Bigg]\,\,.
\label{lagrangian-fluctuations}
\eear
Let us now write the equations of motion  derived from this lagrangian density for the  different fluctuations. First of all, we write the one corresponding to the fluctuations $\chi$ of the cartesian coordinates, namely:
\beq
\partial_m\,\Big[\,r^2\sqrt{\tilde g}\,\,{\cal G}^{mn}\,\partial_n\,\chi\,\Big]\,=\,0\,\,.
\label{chi-eq}
\eeq
The equations for the scalar $\xi$ and the gauge field are coupled. The equation for
$\xi$ is:
\beq
{R^3\over 4k}\,\,
{1\over \sqrt{\tilde g}}\,\partial_m\,\Big[\sqrt{\tilde g}\,
{\cal G}^{mn}\,\partial_n\xi\,\Big]\,+\,
{3\over \sin^2\bar\alpha_n}\,\xi\,+\,{12k\over R^3}\,\,{f_{0r}\over  \sin^3\alpha_n}\,=\,0\,\,,
\label{xi-eq}
\eeq
whereas the equation for the gauge field is:
\beq
{1\over \sqrt{\tilde g}}\,\partial_m\,\Big[\sqrt{\tilde g}\,
{\cal G}^{mp}\,{\cal G}^{nq}\,f_{pq}\,\Big]\,+\,{12k\over R^3}\,\,
{1\over  \sin^3\alpha_n}\,\Big[
\partial_r\xi\,\delta^{n}_{0}\,-\,\partial_0\xi\,\delta^{n}_{r}\,\Big]\,=\,0\,\,.
\label{a-eq}
\eeq

\subsection{Fluctuation of the cartesian coordinates}
Let us consider in detail the fluctuations $\chi$  of the cartesian coordinates. By using the expression of ${\cal G}^{mn}$ in (\ref{inverse-open-metric}), eq. (\ref{chi-eq}) becomes:
\beq
\partial_r\,\big(\,r^4\,\partial_r\,\chi\,\big)\,-\,\partial_0^2\,\chi\,+\,r^2\,\sin^2\alpha_n\,\,
\nabla^2_{\tilde T^{1,1}}\,\chi\,=\,0\,\,.
\eeq
Let us separate variables in this equation as follows:
\beq
\chi\,=\,e^{iEt}\,Y_{l_1,l_2}(\tilde T^{1,1})\,\phi(r)\,\,,
\label{sep-var-cartesian}
\eeq
where $Y_{l_1,l_2}(\tilde T^{1,1})$ are the scalar harmonic functions  for the squashed $T^{1,1}$ (see appendix \ref{harmonics}), which depend, among other quantum numbers, on two integers $l_1$ and $l_2$.  As shown in appendix \ref{harmonics}, the eigenvalues of the laplacian in a squashed $T^{1,1}$ space with general squashing factors depend on $l_1$, $l_2$ and on the R-charge quantum number $r$ (see (\ref{H0-general-squashing})). Remarkably, for the particular squashing of (\ref{squashedT11-n}) the dependence on the R-charge $r$ drops (see (\ref{H0-particular-squashing})). Actually,  by using the eigenvalues of the laplacian derived in the appendix \ref{harmonics}, we get:
\beq
\sin^2\alpha_n\,\,
\nabla^2_{\tilde T^{1,1}}\,Y_{l_1,l_2}\,=\,-\big[\,
l_1(l_1+2)\nu\,+\,l_2(l_2+2)\,(1-\nu)\,\big]\,
Y_{l_1,l_2}\,\,,
\label{laplacian-eigenvalues}
\eeq
with $\nu$ being the filling fraction ($\nu=n/N$). Using this result, the equation for the radial function $\phi(r)$ becomes:
\beq
\partial_r\,\big(\,r^4\,\partial_r\,\phi\,\big)\,+\,E^2\,\phi\,-\,
r^2\,\big[\,l_1(l_1+2)\nu\,+\,l_2(l_2+2)\,(1-\nu)\,\big]\,\phi\,=\,0\,\,.
\eeq
Let us now study the behavior of the solutions of this equation for large $r$. We assume that $\phi\sim r^{\alpha}$ for large $r$. By keeping only the leading terms we get that $\alpha$ must satisfy the following quadratic equation:
\beq
\alpha(\alpha+3)\,=\,l_1(l_1+2)\nu\,+\,l_2(l_2+2)\,(1-\nu)\,\,,
\eeq
which has two solutions, namely:
\beq
\alpha_{\pm}\,=\,-{3\over 2}\,\pm\,
\sqrt{\Big({3\over 2}\Big)^2\,+\,
l_1(l_1+2)\nu\,+\,l_2(l_2+2)\,(1-\nu)}\,\,.
\eeq
Thus, at large $r$, the fluctuation $\phi$ becomes:
\beq
\phi(r)\sim \,c_+\,r^{\alpha_{+}}\,+\,c_-\,r^{\alpha_{-}}\,\,.
\eeq
In general, in $AdS_2$, if a fluctuation behaves in the UV as:
\beq
\chi\sim c_1\,r^{-2a_1}\,+\,c_2\,r^{-2a_2}\,\,,
\qquad\qquad
a_2>a_1\,\,,
\label{UV-exponents}
\eeq
then, the dimension of the operator dual to the normalizable mode is:
\beq
\Delta\,=\,{1\over 2}\,+\,a_2-a_1\,\,.
\label{Delta-a12}
\eeq
In our case:
\beq
a_1\,=\,-{\alpha_+\over 2}\,\,,
\qquad\qquad
a_2\,=\,-{\alpha_-\over 2}\,\,,
\label{a12-alpha-pm}
\eeq
and, therefore, the conformal dimension $\Delta$ is just:
\beq
\Delta\,=\,{1\over 2}\,+\,
\sqrt{\Big({3\over 2}\Big)^2\,+\,
l_1(l_1+2)\nu\,+\,l_2(l_2+2)\,(1-\nu)}\,\,.
\label{Delta-cartesian}
\eeq
Notice that $\Delta=2$ for $l_1=l_2=0$ and, that in general the dimensions are not rational numbers. This is, actually, what happens for the supergravity backgrounds that contain the $T^{1,1}$ space (\ie\ the conifold-like theories). Moreover, the dimensions depend on the filling fraction $\nu$. Only in the symmetric representations in which $l_1=l_2=l$ this dependence on $\nu$ disappears. Indeed, in this case one gets the simple formula:
\beq
\Delta\,=\,{1\over 2}\,+\,
\sqrt{\Big({3\over 2}\Big)^2\,+\,l(l+2)}\,\,,
\qquad\qquad (l_1=l_2=l).
\eeq

\subsection{Coupled modes}

Let us now consider a solution of the equations of motion of the fluctuations in which the gauge field components $a_r$ and $a_i$ are non-zero and given by the ansatz:
\beq
a_r\,=\,e^{iEt}\,Y_{l_1,l_2}(\tilde T^{1,1})\,\phi(r)\,\,,
\qquad\qquad
a_i\,=\,e^{iEt}\,\nabla_i Y_{l_1,l_2}(\tilde T^{1,1})\tilde\phi(r)\,\,.
\eeq
In addition, we will adopt the ansatz in which the scalar field $\xi$ is also fluctuating and given by:
\beq
\xi\,=\,e^{iEt}\,Y_{l_1,l_2}(\tilde T^{1,1})\,z(r)\,\,.
\eeq
For this ansatz of the gauge potential,  the non-vanishing components of the gauge field strength $f=da$  are:
\beq
f_{0r}\,=\,iE\,a_r\,\,,
\qquad
f_{0i}\,=\,iE\,a_i\,\,,
\qquad
f_{ri}\,=\,e^{iEt}\,(\partial_r\tilde\phi-\phi)\,\nabla_i \,
Y_{l_1,l_2}(\tilde T^{1,1})
\,\,.
\qquad
\eeq
We can now write the  different components of the  coupled equations of motion  (\ref{xi-eq}) and  (\ref{a-eq}).  The equation of motion (\ref{a-eq}) for the gauge field for  the index $n=0$ is:
\beq
c_{l_1,l_2}^{\nu}\,\tilde\phi\,=\,r^2\,\partial_r\,\phi\,-\,{3i\over E}\,\,{R^3\over 4k}\,
r^2\,\sin\alpha_n\,\partial_r\,z\,\,.
\label{typeIII-n0}
\eeq
where $c_{l_1,l_2}^{\nu}$ is the eigenvalue of $\sin^2\alpha_n\,\,
\nabla^2_{\tilde T^{1,1}}\,Y_{l_1,l_2}$ (see (\ref{laplacian-eigenvalues})), namely:
\beq
c_{l_1,l_2}^{\nu}\,\equiv\,
l_1(l_1+2)\nu\,+\,l_2(l_2+2)\,(1-\nu)\,\,.
\label{c_l1_l2}
\eeq
For the index $n=r$ the gauge field  equation  (\ref{a-eq})  gives:
\beq
E^2\,\phi\,+\,c_{l_1,l_2}^{\nu}\,r^2\,(\partial_r\tilde\phi-\phi)\,-\,
{3R^3\over 4k}\,\sin\alpha_n\,iE\,z\,=\,0\,\,.
\label{typeIII-nr}
\eeq
Finally, for $n=i$ we get the following relation between $\tilde\phi$ and $\phi$:
\beq
E^2\,\tilde\phi\,=\,-r^2\,\partial_r\,\big[\,r^2\,(\partial_r\,\tilde\phi\,-\,\phi)\,\big]\,\,.
\label{typeIII-ni}
\eeq
One can verify that this last equation (\ref{typeIII-ni}) is a consequence of (\ref{typeIII-n0}) and (\ref{typeIII-nr}). Moreover, we can use (\ref{typeIII-n0}) to eliminate $\tilde\phi$ from the equations of motion of the gauge field. The resulting equation for $\phi$ is:
\beq
\partial_r\,(r^2\partial_r\phi)\,+\,{E^2\over r^2}\,\phi\,-\,
c_{l_1,l_2}^{\nu}\,\phi\,-\,
{3i\over E}\,{R^2\over 4k}\,\sin\alpha_n\,\big[
\partial_r\,(r^2\partial_r\, z)\,+\,{E^2\over r^2}\,z\,\big]\,=\,0\,\,.
\label{typeIII-phi}
\eeq
The equation of motion (\ref{xi-eq}) for the scalar $\xi$ becomes:
\beq
\partial_r\,(r^2\partial_r\, z)\,+\,{E^2\over r^2}\,z\,+\,
(\,3\,-\,c_{l_1,l_2}^{\nu}\,)\,z\,+\,{3iE\over \sin\alpha_n}\,{4k\over R^3}\,\phi\,=\,0\,\,.
\label{typeIII-z}
\eeq
Let us now decouple the two equations (\ref{typeIII-phi}) and (\ref{typeIII-z}). First of all, we define the differential operator $\hat {\cal O}$ as the one that acts on an arbitrary function $\psi$ as:
\beq
\hat {\cal O}\,\psi\,\equiv\,\partial_r\,(r^2\partial_r\, \psi)\,+\,{E^2\over r^2}\,\psi\,\,.
\eeq
Next, we define the functions $\hat z$ and  $\eta$ as:
\beq
\hat z\,=\,-{i\over E}\,{R^3\over 4k}\,\,\sin\alpha_n\,z\,\,,
\qquad\qquad
\eta\,=\,\phi\,-\,{3i\over E}\,{R^3\over 4k}\,\sin\alpha_n\,z\,\,.
\eeq
Then, one can show that (\ref{typeIII-phi}) and (\ref{typeIII-z}) can be written as the following matrix equation:
\beq
\Big(\,\,\hat{\cal O}\,-\,{\cal M}\,\,\Big)\,
\left(\begin{array}{c}
\,\hat z\\
\eta
\end{array}\right)
=\,0\,\,,
\label{matrix-fluct-eq}
\eeq
where ${\cal M}$ is the following $2\times 2$ matrix:
\beq
{\cal M}\,=\,\left(
\begin{array}{ccc}
c_{l_1,l_2}^{\nu}+6&&-3\\\\\\
-3c_{l_1,l_2}^{\nu}&&c_{l_1,l_2}^{\nu}
\end{array}
\right)\,\,.
\label{M-matrix}
\eeq

In what follows we will distinguish the case in which $c_{l_1,l_2}^{\nu}\not=0$ from the case in which it vanishes and the fluctuation is an s-wave in the internal $\tilde T^{1,1}$ space with $l_1=l_2=0$. Actually, it follows  from (\ref{typeIII-n0}) and (\ref{typeIII-nr}) that $\eta=0$ when $c_{l_1,l_2}^{\nu}=0$ and, therefore, in this case there is only one type of mode, namely that corresponding to $\hat z$. Notice that (\ref{matrix-fluct-eq}) is consistent with having $\eta=0$ when $c_{l_1,l_2}^{\nu}=0$. 

Let us consider from now on the case $c_{l_1,l_2}^{\nu}\not=0$. In this case we have to deal with the full matrix equation (\ref{matrix-fluct-eq}).  The matrix ${\cal M}$ written above is diagonalizable and its two eigenvalues are:
\beq
\lambda_{l_1,l_2}^{(\pm)}\,=\,c_{l_1,l_2}^{\nu}\,+\,3\,\big(\,1\,\pm
\sqrt{1+c_{l_1,l_2}^{\nu}}\,\big)\,\,.
\label{typeIII-eigenvalues}
\eeq
The corresponding eigenfunctions are the following combinations of $\hat z$ and $\eta$:
\beq
\psi_{\pm}\,=\,\big(\,\sqrt{1+c_{l_1,l_2}^{\nu}}\,\pm 1\,\big)\,\hat z\,\mp\,\eta\,\,.
\eeq
Indeed, one can check that $\psi_{\pm}$ satisfy the following decoupled equations:
\beq
\partial_r\,(r^2\partial_r\, \psi_{\pm})\,+\,\Big(\,{E^2\over r^2}\,-\,
\lambda_{l_1,l_2}^{(\pm)}\,\Big)\,\psi_{\pm}\,=\,0\,\,.
\label{typeIII-decoupled}
\eeq
In order to obtain the conformal dimensions associated to the normalizable solutions of (\ref{typeIII-decoupled}), let us study the behavior of the functions $\psi_{\pm}(r)$ for  large $r$. By considering functions of the type $\psi_{\pm}(r)\sim r^a$, one gets two possible solutions for $a$ and, thus, $\psi_{\pm}(r)$ behaves for large $r$ as:
\beq
\psi_{\pm}(r)\sim c_1\,r^{-2a_1^{(\pm)}}\,+\,c_2\,r^{-2a_2^{(\pm)}}\,\,,
\eeq
where $c_1$ and $c_2$ are constants and the exponents $a_1^{(\pm)}$ and 
$a_2^{(\pm)}$ are given by:
\beq
a_1^{(\pm)}\,=\,{1-\sqrt{1+4 \,\lambda_{l_1,l_2}^{(\pm)}}\over 4}\,\,,
\qquad\qquad
a_2^{(\pm)}\,=\,{1+\sqrt{1+4 \,\lambda_{l_1,l_2}^{(\pm)}}\over 4}\,\,.
\eeq
The conformal dimensions $\Delta_{\pm}$ of the operators dual to the normalizable  modes of $\psi_{\pm}(r)$ are immediately obtained from the general equation (\ref{Delta-a12}):
\beq
\Delta_{\pm}\,=\,{1\over 2}\,\,\Big(\,1\,+\,\sqrt{1+4\,\lambda_{l_1,l_2}^{(\pm)}}\,\Big)\,\,.
\eeq
Remarkably, $1+4\,\lambda_{l_1,l_2}^{(\pm)}$ is a perfect square. Indeed, by using (\ref{typeIII-eigenvalues}) one can straightforwardly  prove that:
\beq
1+4\,\lambda_{l_1,l_2}^{(\pm)}\,=\,\Big(\,
2\,\sqrt{1+c_{l_1, l_2}^{\nu}}\pm 3\,\Big)^2\,\,.
\eeq
Therefore, the conformal dimensions  $\Delta_{\pm}$  can be simply be written as:
\beq
\Delta_{+}\,=\,2\,+\,\sqrt{1+c_{l_1, l_2}^{\nu}}\,\,,
\qquad\qquad
\Delta_{-}\,=\,\sqrt{1+c_{l_1, l_2}^{\nu}}\,-\,1\,\,.
\label{dimensions-Delta-pm}
\eeq
Recall that, to obtain (\ref{dimensions-Delta-pm}), we have assumed that 
$c_{l_1, l_2}^{\nu}>0$, which ensures that $\Delta_{-}>0$. When $c_{l_1, l_2}^{\nu}$ vanishes then $\eta=0$ and the matrix equation (\ref{matrix-fluct-eq}) reduces to the simple equation $\big(\,\hat {\cal O}\,-\,6\,\big)\,\hat z\,=\,0$. It is then straightforward to demonstrate that the normalizable mode has conformal dimension $\Delta=3$, which is the same value as the one that is obtained by substituting  $c_{l_1, l_2}^{\nu}=0$ in the equation giving $\Delta_+$ in (\ref{dimensions-Delta-pm}).

The dimensions (\ref{dimensions-Delta-pm}) are, in general, not rational and they depend on the filling fraction $\nu$. However, for symmetric representations in which $l_1=l_2=l$ the dependence on $\nu$ disappears and one has:
\beq
c_{l,l}^{\nu}\,=\,l(l+2)\,\,
\Longrightarrow
\sqrt{1+c_{l, l}^{\nu}}\,=\,l+1\,\,.
\eeq
Therefore, in this symmetric case the dimensions are rational and are given by:
\beq
\Delta_{+}\,=\,3\,+l\,\,,\qquad (l\ge 0)\,\,,
\qquad\qquad
\Delta_{-}\,=\,l\,\,,\qquad (l>0)\,\,,
\eeq
where we have already taken into account the $l=0$ case.

\subsection{Internal gauge field modes}

We now consider fluctuation modes of the gauge field in which the only components which are non-vanishing are those along the directions of the internal $\tilde T^{1,1}$ space (they will be denoted by $a_i$). In this case the non-zero components of the field strength $f$ are:
\beq
f_{0i}\,=\,\partial_0\,a_i\,\,,\qquad\qquad
f_{ri}\,=\,\partial_r\,a_i\,\,,\qquad\qquad
f_{ij}\,=\,\partial_i\,a_j\,-\,\partial_j\,a_i\,\,.
\eeq
The equations of motion (\ref{a-eq}) for the gauge field for $n=0, r$ are just:
\beq
\partial_{0}\big(\,\nabla^{i}\,a_i\,\big)\,=\,0\,\,,\qquad\qquad
\partial_{r}\big(\,\nabla^{i}\,a_i\,\big)\,=\,0\,\,,
\eeq
where:
\beq
\nabla^{i}\,a_i\,=\,
{1\over \sqrt{\tilde g}}\,\partial_i\,\big(\, \sqrt{\tilde g}\,
\tilde g^{ij}\,a_j\,\big)\,\,.
\eeq
Thus,  clearly we should require the fulfillment of the following transversality condition:
\beq
\nabla^{i}\,a_i\,=\,0\,\,.
\label{nabla-a}
\eeq
We now write the equation  (\ref{a-eq})  for $n=i$, namely:
\beq
\tilde g^{ij}\,\Big[\,\partial_r\,\big[\,r^{2}\,\partial_r\,a_j\,\big]\,-\,
\,r^{-2}\,\partial_{0}^2\,a_j\,\Big]\,+\,
{\sin^2\alpha_n\over \sqrt{\tilde g}}\,\partial_k\,\big[\sqrt{\tilde g}\,(
\partial^k a^i-\partial^i a^k\,)\,\big]\,=\,0\,\,.
\label{eom-ai}
\eeq
The last term in this equation can be rewritten as:
\beq
{1\over \sqrt{\tilde g}}\,\partial_k\,\big[\sqrt{\tilde g}\,(
\partial^k a^i-\partial^i a^k\,)\,\big]\,=\,\nabla_k\nabla^k\,a^{i}\,-\,
R^{i}_{\,\,k}\,a^k\,-\,\nabla^{i}\,\nabla_k\,a^k\,\,,
\label{df}
\eeq
where $R^{i}_{\,\,k}$ is the Ricci tensor of the $\tilde T^{1,1}$ space. Taking (\ref{nabla-a}) into account, the last term in  (\ref{df}) vanishes and the right-hand side of (\ref{df})
can be written in terms of the Hodge-de Rham operator $\Delta_1$, which acts on a vector field with components $f_i$ as:
\beq
\Delta_1\,f_i\,\equiv\, -\nabla_k\nabla^k\,f_{i}\,+\,
R^{k}_{i}\,f_k\,\,.
\label{HdR}
\eeq
Thus, the equation of motion (\ref{eom-ai}) becomes:
\beq
\partial_r\,\big[\,r^{2}\,\partial_r\,a_i\,\big]\,-\,
\,{1\over r^2}\,\partial_{0}^2\,a_i\,-\,4\nu\,(1-\nu)\,\Delta_1\,a_i\,=\,0\,\,,
\label{fluct-eq-ai}
\eeq
where we have taken into account that $\sin^2\alpha_n\,=\,4\nu\,(1-\nu)$. 
Let us next  separate variables  in $a_i$ as:
\beq
a_i\,=\,e^{iEt}\,Y_{i}(\tilde T^{1,1})\,\phi(r)\,\,,
\label{typeI-sep-var}
\eeq
where $Y_{i}(\tilde T^{1,1})$ is a vector spherical harmonic for the 
$\tilde T^{1,1}$ space which satisfies the transversality condition:
\beq
\nabla^{i}\,Y_{i}(\tilde T^{1,1})\,=\,0\,\,.
\eeq
Let us diagonalize the Hodge-de Rham operator $\Delta_1$ in the space of the
$\tilde T^{1,1}$ vector harmonics and  write:
\beq
4\nu \,(1-\nu)\,\Delta_1\,Y_{i}\,=\,\Lambda\,Y_{i}\,\,,
\label{Delta1-Lambda}
\eeq
where the eigenvalue $\Lambda$ will depend on the quantum numbers $l_1$ and $l_2$, as well as  on the filling fraction $\nu$ (see below). By plugging the ansatz (\ref{typeI-sep-var}) on the fluctuation equation (\ref{fluct-eq-ai}), one gets the following equation for the radial function $\phi(r)$:
\beq
\partial_r\,\big[\,r^{2}\,\partial_r\,\phi\,\big]\,+\,{E^2\over r^2}\,\phi
\,-\,\Lambda\,\phi\,=\,0\,\,.
\eeq
This equation can be solved for large $r$ by taking $\phi\sim r^{\alpha}$, with $\alpha$ being some exponent. Taking into account that the term with the energy is subleading at large $r$, one gets that the exponent $\alpha$ must satisfy the following quadratic equation:
\beq
\alpha\,(\alpha+1\,)\,=\,\Lambda\,\,,
\eeq
which has two solutions, namely:
\beq
\alpha_{\pm}\,=\,{-1\pm\sqrt{1+4\Lambda}\over 2}\,\,.
\eeq
Thus, $\phi(r)$ behaves as (\ref{UV-exponents}) with 
$a_1=-\alpha_+/2$ and $a_2=-\alpha_-/2$, \ie\ as in (\ref{a12-alpha-pm}). Therefore, the conformal dimension of the operator dual to these fluctuations is:
\beq
\Delta\,=\,{1\over 2}\,\,\Big(\,1\,+\,\sqrt{1+4\Lambda}\,\Big)\,\,.
\eeq
Let us analyze the different dimensions obtained for the various vector harmonics for the $\tilde T^{1,1}$ space that were studied in appendix \ref{harmonic}. As shown in this appendix,  the results depend on whether the filling fraction $\nu$ is arbitrary or $\nu=1/2$. These two cases will be studied separately in what follows.

\subsubsection{Arbitrary filling}
As argued in  appendix \ref{harmonic}  in this case we should take vanishing R-charge $r$ and the eigenvalues $\Lambda$ depend on the remaining two quantum numbers $l_1$ and $l_2$. Actually,  there are two series of eigenvalues, which were denoted by $\lambda_1$ and $\lambda_2$ in the appendix. Let us consider the first of these modes and let us denote by $\Lambda_1$ the corresponding value of $\Lambda$ in (\ref{Delta1-Lambda}). From the value of $\lambda_1$ written in (\ref{lambda12}), we get:
\beq
\Lambda_1\,=\,c_{l_1, l_2}^{\nu}\,\,.
\eeq
The associated conformal dimension $\Delta_1$ is given by:
\beq
\Delta_1\,=\,{1\over 2}\,\,\Big(\,1\,+\,\sqrt{1+4\,c_{l_1, l_2}^{\nu}}\,\Big)\,\,,
\label{Delta1}
\eeq
which is, in general, not rational. Moreover, only for $l_1=l_2=l$ the dimension $\Delta_1$ is independent of the filling fraction since 
$c^{\nu}_{l,l}=l(l+1)$.

The second class of eigenvalues in this arbitrary filling case, denoted by $\lambda_2^{(\pm)}$ in (\ref{lambda12}), lead to a value of $\Lambda$ given by:
\beq
\Lambda_2^{(\pm)}\,=\,1\,+\,c_{l_1, l_2}^{\nu}\pm
\sqrt{1+4\nu^3\, l_1(l_1+2)\,+\,4(1-\nu)^3\,l_2(l_2+2)}\,\,.
\eeq
The corresponding conformal dimension is:
\beq
\Delta_2^{(\pm)}\,=\,{1\over 2}\,+\,
\sqrt{{5\over 4}\,+\,c_{l_1, l_2}^{\nu}\pm
\sqrt{1+4\nu^3\, l_1(l_1+2)\,+\,4(1-\nu)^3\,l_2(l_2+2)}
}\,\,.
\label{Delta2}
\eeq
We will see below that this formula simplifies greatly if $\nu=1/2$.

\subsubsection{Half filling}

When $\nu=1/2$,  it was shown in appendix \ref{harmonic} that there are transverse modes for $r\not=0$. As in the generic $\nu$  case, there are two types of modes. The eigenvalues of the first type were denoted by $\tilde\lambda_1$ in (\ref{tilde-lambda12}) and they lead to the following eigenvalue in (\ref{Delta1-Lambda}):
\beq
\tilde \Lambda_1^{\pm}\,=\,\tilde c_{l_1,l_2}\pm r\,\,,
\eeq
with $\tilde c_{l_1,l_2}$ being the value of $c_{l_1, l_2}^{\nu}$ for 
$\nu=1/2$, \ie:
\beq
\tilde c_{l_1,l_2}\,\equiv\,c_{l_1, l_2}^{\nu=1/2}\,=\,
{l_1(l_1+2)\over 2}\,+\,{l_1(l_1+2)\over 2}\,\,.
\eeq
The dimension of these modes are non-rational and given by:
\beq
\tilde\Delta_1^{(\pm)}\,=\,
{1\over 2}\,+\,\sqrt{{1\over 4}\,+\,
\tilde c_{l_1,l_2}\pm r}\,\,.
\label{tilde-Delta1}
\eeq
For $r=0$ these dimensions reduce to the ones written in (\ref{Delta1}) for $\nu=1/2$. 

Next, let us consider the modes with eigenvalues $\tilde \lambda_2^{(\pm)}$, for which the eigenvalue defined in (\ref{Delta1-Lambda}) is independent of $r$ and given by:
\beq
\tilde\Lambda_2^{(\pm)}\,=\,1\,+\,\tilde c_{l_1,l_2}\pm
\sqrt{1+\tilde c_{l_1,l_2}}\,\,.
\eeq
Remarkably, one can verify that $1+4\, \tilde\Lambda_2^{(\pm)}$ can be written as a square.  Indeed, one can straightforwardly check that:
\beq
1+4\, \tilde\Lambda_2^{(\pm)}\,=\,\Big(\,2\,\sqrt{1+\tilde c_{l_1,l_2}}\pm 1\,\Big)^2\,\,.
\eeq
Thus, the associated conformal dimensions $\tilde\Delta_2^{(+)}$ and $\tilde\Delta_2^{(-)}$ are just:
\beq
\tilde\Delta_2^{(+)}\,=\,1\,+\,\sqrt{1+\tilde c_{l_1,l_2}}\,\,,
\qquad\qquad
\tilde\Delta_2^{(-)}\,=\,\sqrt{1+\tilde c_{l_1,l_2}}\,\,.
\eeq
These dimensions are generically not rational. However, 
when $l_1=l_2=l$, one has that
$\sqrt c_{l,l}=l(l+2)$ and, thus,  $\sqrt{1+c_{l,l}}=l+1$. Therefore, in this  $l_1=l_2=l$ case 
the dimensions associated to  these modes are rational and given by:
\beq
\tilde\Delta_2^{(+)}\,=\,2+l\,\,,\qquad\qquad
\tilde\Delta_2^{(-)}\,=\,1+l\,\,,
\qquad\qquad (l_1=l_2=l)\,\,.
\eeq

\section{Dimer fluctuations}
\label{dimer-fluct}

Let us now consider the fluctuations around a connected configuration of the type studied in section \ref{dimers}. We will restrict ourselves to analyze the case in which the background is supersymmetric, \ie\ for vanishing temperature. Recall that in this case the D6-branes do not reach the origin at $r=0$ and they extend along the Minkowski directions (say, along the coordinate $x$). For this reason we will have to distinguish between longitudinal and transverse fluctuations in the $(x,y)$ plane. By computing the lagrangian for generic quadratic fluctuations one realizes that most of the modes are coupled and the diagonalization of the equations of motion is rather involved. However, one can verify that the fluctuation of the transverse position in the Minkowski space is decoupled from the other modes. In this paper we will limit ourselves to studying these transverse fluctuations. The corresponding lagrangian density and equation of motion are derived in appendix \ref{fluctuations} (eqs. (\ref{dimer-fluct-lagrangian}) and (\ref{chi-eq-hanging}), respectively). More explicitly, the equation of motion for the transverse modes can be written as:
\beq
\partial_r\,\big(\,r^2\, \sqrt{r^4-r_*^4}\,
\,\partial_r\,\chi\,\big)\,-\,{r^2\over \sqrt{r^4-r_*^4}}\,\partial_0^2\,\chi\,+\,
{r^4\over \sqrt{r^4-r_*^4}}\,\sin^2\alpha_n\,\,
\nabla^2_{\tilde T^{1,1}}\,\chi\,=\,0\,\,,
\label{transverse-fluct-dimer}
\eeq
where $r_*$ is the minimal value of the $r$ coordinate attained by the unperturbed configuration of the  brane. Notice  also that any value of $r$ in the interval $[r_*, +\infty)$ is reached twice by the brane.

Let us separate variables in (\ref{transverse-fluct-dimer}) as in (\ref{sep-var-cartesian}) and let us define the rescaled radial coordinate $\rho$ and the rescaled energy $\bar E$ as:
\beq
\rho\,=\,{r\over r_*}\,\,,\qquad\qquad
\bar E\,=\,{E\over r_*}\,\,.
\label{rho-barE}
\eeq
In terms of these quantities, the equation of motion for the  fluctuation becomes:
\beq
\partial_{\rho}\,\big(\,\rho^2\, \sqrt{\rho^4-1}\,
\,\partial_{\rho}\,\phi\,\big)\,+\,\bar E^2\,
{\rho^2\over \sqrt{\rho^4-1}}\,\phi\,-\,c_{l_1,l_2}^{\nu}
{\rho^4\over \sqrt{\rho^4-1}}\,\phi\,=\,0\,\,,
\label{fluct-eq-cartesian-hanging}
\eeq
where $c_{l_1,l_2}^{\nu}$ has been defined in (\ref{c_l1_l2}). 

When $c_{l_1,l_2}^{\nu}=0$ (or equivalently for the s-wave modes with $l_1=l_2=0$), the fluctuation equation (\ref{fluct-eq-cartesian-hanging}) can be exactly solved by means of a change of variables found in \cite{Klebanov:2006jj, Brower:2006hf}. Indeed,  let us introduce a new variable $\eta$, related to $\rho$ by means of the equation:
\beq
\eta\,=\,\int_{1}^{\rho}\,
{d\bar\rho\over (\bar\rho^2+\bar E^2)\sqrt{\bar\rho^4-1}}\,\,.
\eeq
By this change of variables $\rho \in [1,+\infty]$ is mapped into $\eta\in [0,\bar \eta(\bar E)\,]$, where $\bar \eta(\bar E)$ is defined as:
\beq
\bar \eta(\bar E)\,\equiv\,\int_{1}^{\infty}\,
{d\bar\rho\over (\bar\rho^2+\bar E^2)\sqrt{\bar\rho^4-1}}\,\,.
\eeq
Next, we introduce a new function $\Phi(\eta)$ as follows:
\beq
\Phi(\eta)\,=\,{\rho\over \sqrt{\rho^2+\bar E^2}}\,\phi(\rho)\,\,.
\eeq
By explicit calculation one can prove that the equation satisfied by $\Phi(\eta)$ is simply:
\beq
{d^2\Phi\over d \eta^2}\,+\,{1\over 4}\,\bar E^2\,(\bar E^4-1)\,
\Phi\,=\,0\,\,,
\eeq
which can be immediately integrated in terms of trigonometric functions. By imposing that $\Phi$ vanishes at $\eta=0, \bar \eta(\bar E)$, we get the following quantization condition on the energy $\bar E$:
\beq
\bar E\,\sqrt{\bar E^4-1}\,\,\bar \eta(\bar E)\,=\,n\pi\,\,,
\qquad\qquad
n=1,2,\cdots\,.
\eeq
This equation must be solved numerically in order to get the energy levels.

Let us come back to the analysis of the equation (\ref{fluct-eq-cartesian-hanging}) for general values of the Kaluza-Klein quantum numbers $l_1$ and $l_2$. We will show that this equation can be mapped to the Schr\"odinger equation of a quantum integrable model, namely the so-called Inozemtsev $BC_1$ model. The first step to prove this result is changing  the radial variable $\rho$ by a new variable $z$, defined as:
\beq
z\,=\,1-\rho^2\,\,.
\eeq
Then,  one can check that the fluctuation equation (\ref{fluct-eq-cartesian-hanging}) can be written as:
\beq
{d^2\phi\over dz^2}\,+\,{1\over 2}\,\,
\Big[\,{1\over z}\,+\,{3\over z-1}\,+\,{1\over z-2}\,\Big]\,\partial_z\,\phi\,-\,
{1\over 4}\,\,{c_{l_1,l_2}^{\nu}(z-1)\,+\,\bar E^2\over z(z-1)(z-2)}\,\phi\,=\,0\,\,.
\label{heun-cartesian-fluct}
\eeq
This equation is a Heun's equation, whose more general expression is  given in appendix 
\ref{Heun} (eq. (\ref{Heun-eq})). As reviewed in appendix \ref{Heun}, the Heun equation is an ordinary differential equation with four regular singular points whose location and characteristic exponents are parametrized by several numbers, which we denoted by 
$\alpha$, $\beta$, $\gamma$, $\delta$, $\epsilon$, $q$ and $d$ (see (\ref{Heun-eq})). 
The solution of the general Heun equation (\ref{Heun-eq}) which is regular at $z=0$  is denoted 
$Hl(d,q;\alpha,\beta,\gamma,\delta;z)$. 

By direct comparison of (\ref{heun-cartesian-fluct}) and (\ref{Heun-eq}), one gets that in our case $d=2$ and that 
the coefficients $\gamma$, $\delta$, $\epsilon$ and $q$ are:
\beq
\gamma\,=\,{1\over 2}\,\,,\qquad\qquad
\delta\,=\,{3\over 2}\,\,,\qquad\qquad
\epsilon\,=\,{1\over 2}\,\,\qquad\qquad
q\,=\,-{c_{l_1,l_2}^{\nu}\over 4}\,+\,{\bar E^2\over 4}\,\,.
\eeq
Notice also that the remaining parameters $\alpha$ and $\beta$ should satisfy (see (\ref{Heun-condition})):
\beq
\alpha\,+\,\beta\,=\,\gamma+\delta\,+\,\epsilon\,-\,1\,=\,{3\over 2}\,\,,
\qquad\qquad
\alpha\beta\,=-\,{c_{l_1,l_2}^{\nu}\over 4}\,\,.
\eeq
In order to determine $\alpha$ and $\beta$ we use the fact that  that they are the roots of the quadratic equation:
\beq
x^2\,-\,(\alpha+\beta)\,x\,+\alpha\beta\,=\,x^2\,-\,{3\over 2}\,x\,
-\,{c_{l_1,l_2}^{\nu}\over 4}\,=0\,\,,
\eeq
which are:
\beq
\alpha\,=\,{\Delta+1\over 2}\,\,,\qquad\qquad
\beta\,=\,{2-\Delta\over 2}\,\,,
\eeq
with $\Delta$ being the conformal dimension (\ref{Delta-cartesian}). Then, the solution of the fluctuation equation which is regular at $\rho=1$ is:
\beq
\phi(\rho)\,=\,
Hl\big(2,{\bar E^2+(\Delta+1)(2-\Delta)\over 4};{\Delta+1\over 2}, {2-\Delta\over 2},
{1\over 2}, {3\over 2};1-\rho^2\big)\,\,.
\eeq
When $\bar E=0$ we can rewrite this result in terms of a hypergeometric function (see (\ref{heun-hyper})), namely:
\beq
\phi(\rho)\big|_{\bar E=0}\,=\,
F\big(\,{\Delta+1\over 4}\,,\,{2-\Delta\over 4}\,,\,{1\over 2}\,;\,1-\rho^4\,\big)\,\,.
\eeq

As reviewed in appendix \ref{Heun}, with an additional change of variable, the Heun equation can be converted into a Schr\"odinger equation for a particle moving under the influence of a potential which is a combination of elliptic functions. For a generic Heun equation this potential was written in eq. (\ref{I-hamiltonian}). For our particular case of (\ref{heun-cartesian-fluct}) the hamiltonian takes the form:
\beq
H\,=\,-{d^2\over dx^2}\,+\,\Delta (\Delta-1)\,\,
\wp (x)\,\,,
\label{I-hamiltonian-cartesian}
\eeq
where $\wp (x)$ is the Weierstrass function for periods $(2\omega_1, 2\omega_3)=(1,i)$ (see eq. (\ref{wp}) for its general definition). The relation between the variable $x$ in (\ref{I-hamiltonian-cartesian}) and the coordinate $z$ of the Heun equation has been written in (\ref{z-x-d2}). It is interesting to write here the relation between our original radial variable $r$ and the new variable $x$. One gets:
\beq
r\,=\,{L\,r_*^2\over \pi}\,\,\sqrt{\wp(x)}\,\,,
\eeq
where $L$ is the length  of the dimer (see  eq. (\ref{L-dimer-SUSY-explicit})).  By means of this change of variable the equation for the fluctuation can be written as:
\beq
H\,\phi\,=\,{\cal E}\,\phi\,\,,
\label{Sch}
\eeq
where $\phi$ is the same function as in (\ref{heun-cartesian-fluct}) but now considered as a function of $x$. Moreover, the energy ${\cal E}$ of the Schr\"odinger problem (\ref{Sch}) is related to the energy $\bar E$. Indeed, by using (\ref{calE-q-A}) one gets:
\beq
\bar E^2\,=\,{8\pi \over 
\Big[\,\Gamma\big({1\over 4}\big)\Big]^2}\,\,{\cal E}\,\,.
\eeq
Notice that the Schr\"odinger equation (\ref{Sch}) for the hamiltonian  (\ref{I-hamiltonian-cartesian}) is nothing but the Lam\'e equation. When $\Delta\in{\mathbb Z}$ the eigenfunctions and band structure of the Lam\'e equation  have been obtained long ago \cite{Lame}. Notice, however, that in our case 
$\Delta\in{\mathbb Z}$ only when $l_1=l_2=0$ (see (\ref{Delta-cartesian})).

The coordinate $x$ in (\ref{I-hamiltonian-cartesian}) takes values in the interval $[0,1]$, where the Weierstrass function $\wp(x)$ with periods $(2\omega_1, 2\omega_3)=(1,i)$
is real and positive and has a minimum at $x=1/2$. Moreover, $\wp(x)\to +\infty$ when $x\to 0,1$, which means that the eigenvalue problem has a discrete spectrum consisting of an infinite tower of eigenvalues. Although many results are known for these type of integrable models, we will leave this analysis for a future work and here we will content ourselves with estimating the eigenvalues by means of the WKB method. Actually, this estimate can be performed directly in the variables of eq. (\ref{transverse-fluct-dimer}). Indeed, by using the equations written in \cite{RS}, we get that the energy levels can be estimated as:
\beq
\bar E_{n}\,\simeq\,{\pi\over \zeta}\,\,
\sqrt{
\big(\,n+1\,\big)\,\big(\,n+\,\Delta\,)}\,\,,\qquad\qquad
(n\ge 0)\,\,,
\eeq
where $\zeta$ is given by the following integral:
\beq
\zeta\,=\,\int_1^{\infty}\,\,{d\rho\over \sqrt{\rho^4-1}}\,=\,
{1\over 4\sqrt{2\pi}}\,
\Big[\,\Gamma\big({1\over 4}\big)\Big]^2\,\,.
\eeq
Let us write this result in  terms of the length $L$ of the dimer. The relation between $L$ and $r_*$ was written in eq. (\ref{L-dimer-SUSY-explicit}). By using this result, one finds that the un-rescaled  energies $E_n$ are given by:
\beq
 E_{n}\,\simeq\,{16\pi^3\over 
\big[\,\Gamma\big({1\over 4}\big)\big]^4\,L}\,\,
\sqrt{
\big(\,n+1\,\big)\,\big(\,n+\,\Delta\,)}
\,\,,\qquad\qquad
(n\ge 0)\,\,.
\eeq

\section{Summary and conclusions}
\label{conclusions}

Let us briefly recapitulate the main content of this work. We have studied the addition of localized fermionic impurities to the ABJM Chern-Simons-matter model in 2+1 dimensions. In the holographic approach the impurities are added by means of D6-branes extended along the holographic coordinate and wrapping a squashed $T^{1,1}$ space at a fixed point of the Minkowski spacetime.  The coupling of the D6-brane to the RR field of the background induces an electric worldvolume gauge field which prevents the collapse of the wrapped brane and must be quantized accordingly.

The number $n$ of defect fermions fixes the size and the internal deformation of the $T^{1,1}$ space. We have studied the thermodynamics of the system by analyzing the D6-brane action in a black hole background and we have determined the entropy and free energy of the impurity. We have also studied connected configurations in which the branes are hanging from the boundary and we have shown that these configurations dominate over the unconnected ones at low temperature, giving rise to a dimerized phase.

A crucial point in all the approach is the stability of the D6-brane embeddings used to introduce the holographic impurities. In order to study this aspect of the system,  we have performed a detailed analysis of the fluctuations of the D6-brane probe in the ABJM background and we have determined the spectrum of conformal dimensions of the dual operators in the defect theory. This required a detailed analysis of the KK modes of the probe brane and of the harmonic analysis on the squashed $T^{1,1}$ space which uncovered a very rich structure. We also studied a subset of fluctuations of the connected configurations and pointed out a connection with an elliptic quantum  integrable  model.

There are many unanswered questions  raised by our results which could be the subject of future research. First of all, to complete the holographic setup we should have a more detailed description of the field theory side of the correspondence. In particular, we would like to determine the concrete action of the fermionic fields of the impurities and their coupling to the different bulk fields. This would allow us to write the precise dictionary between fluctuations of the probe and impurity operators. Moreover, in order to complete the spectroscopic analysis of section \ref{Fluct} we would have to study the fermionic fluctuations of the D6-brane probe and we should verify how the different modes are accommodated in supermultiplets. 

All our results have been obtained in the probe approximation. As argued in \cite{Harrison:2011fs}, there are some physical effects which are not captured by this approximation and that require taking into account the backreaction from the impurity. In this sense it would be interesting to find a generalization of the bubbling geometries \cite{Lin:2004nb} for the ABJM Wilson lines, similar to the ones  found in \cite{Yamaguchi:2006te,Lunin:2006xr,D'Hoker:2007fq} for the $AdS_5\times S^5$ geometry.

On general grounds it is also of great interest the extension of  our results to less supersymmetric configurations and/or to systems in which the ambient theory is not conformal. In the first case one could study impurity theories constructed from the brane embeddings of 
\cite{Karaiskos:2011kf}.  In the case of non-conformal bulk theories we have studied the addition of impurities to backgrounds generated by Dp-branes, with $p\not= 3$. The results will be presented elsewhere \cite{Kondopaper}.

\section*{Acknowledgments}
 
We are grateful to Eduardo Conde, Wolfgang M\"uck, Diego Rodr\'\i guez-G\'omez,  Konstadinos Sfetsos and Konstadinos  Siampos for useful discussions.  P.B. would also like to thank
the developers of SAGE \cite{sage}, Maxima \cite{maxima}, Numpy and Scipy \cite{scipy}. This work is funded in part by MICINN under grant FPA2008-01838, by the Spanish Consolider-Ingenio 2010 Programme CPAN (CSD2007-00042) and by Xunta de Galicia (Conseller{\'i}a de Educac{\'i}on, grant
INCITE09 206 121 PR and grant PGIDIT10PXIB206075PR) and by FEDER. P.B. is supported as well by the MInisterio de Ciencia e INNovaci{\'on} through the Juan de la Cierva program.

\vskip 1cm
\renewcommand{\theequation}{\rm{A}.\arabic{equation}}
\setcounter{equation}{0}
\medskip
\appendix

\section{BPS configurations}
\label{baryon}

The Euler-Lagrange equation of motion derived from the hamiltonian density (\ref{hamiltonian_density}) is:
\beq
{d\over dr}\,\Bigg[
{r^2\alpha'\over \sqrt{1\,+\,r^2\,\big(\alpha'\big)^2}}\,\,
\sqrt{\big(\sin\alpha\big)^6\,+\,{\cal C}_n(\alpha)^2}\Bigg]\,=\,
3\,{\sqrt{1\,+\,r^2\,\big(\alpha'\big)^2}\over 
\sqrt{\big(\sin\alpha\big)^6\,+\,{\cal C}_n(\alpha)^2}}\,\,
(\sin\alpha )^3\,\,\Lambda_n(\alpha)\,\,,\qquad
\label{embedding-eom}
\eeq
where $\Lambda_n(\alpha)$ is the function defined in (\ref{Lambda_n}). In the main text we have studied solutions of these equations with constant  $\alpha$.  We will now find a more general class of solutions of (\ref{embedding-eom}). Instead of dealing with the Euler-Lagrange equations we will follow a different strategy by establishing a BPS bound for the energy (see \cite{Craps:1999nc,Camino:1999xx}). The configurations that saturate this bound solve the equations of motion and will be characterized by a first-order differential equation which can be integrated in analytic form. We will start  by rewriting the hamiltonian  density (\ref{hamiltonian_density}) as:
\beq
{\cal H}\,=\,{R^9\over 2^{10}\,\pi^3\,k^2}\,
\,\, \sqrt{{\cal Z}^2\,+\,{\cal Y}^2
}\,\,,
\eeq
where ${\cal Z}$ and ${\cal Y}$ are given by:
\bear
&&{\cal Z}\,=\,r\,\alpha'\,
\big[\,(\sin\alpha)^3\,\cos\alpha\,-\,{\cal C}_n(\alpha)\,\sin\alpha\,\big]\,+\,
(\sin\alpha)^4\,+\,{\cal C}_n(\alpha)\,\cos\alpha\,\,,\rc\rc
&&{\cal Y}\,=\,(\sin\alpha)^3\,\cos\alpha\,-\,{\cal C}_n(\alpha)\,\sin\alpha\,-\,
r\,\alpha'\,\big[\,(\sin\alpha)^4\,+\,{\cal C}_n(\alpha)\,\cos\alpha\,\big]\,\,,
\eear
and ${\cal C}_n(\alpha)$ is the function defined in (\ref{calCn}). 
Clearly, the energy functional $H$ satisfies the bound:
\beq
H\,\ge\,{R^9\over 2^{10}\,\pi^3\,k^2}\,
\int dr\,\big|\,{\cal Z}\,\big|\,\,,
\label{bound}
\eeq
which is saturated by the configurations that satisfy ${\cal Y}=0$.  Interestingly,  for any function $\alpha(r)$, the function ${\cal Z}$ can be written as a total derivative:
\beq
{\cal Z}\,=\,{d\over dr}\,\,\Big[\,r\,\Big(\,(\sin\alpha)^2\,-\,\Lambda_n(\alpha)\,\cos\alpha\,\Big)\,\Big]\,\,.
\label{CalZ-derivative}
\eeq
As a consequence of (\ref{CalZ-derivative}),  only the boundary values of $\alpha(r)$ matter when one evaluates the right-hand side of (\ref{bound}) and, thus, the functions $\alpha(r)$ which saturate the bound correspond to D6-brane embeddings which, for given boundary conditions, minimize the energy. Let us now analyze in detail these minimal energy configurations.  In order to write the   BPS condition ${\cal Y}=0$  in a suitable way, let us recast  ${\cal Y}$ in the form:
\beq
{\cal Y}\,=\,\sin\alpha\,\Lambda_n(\alpha)\,-\,r\,\alpha'\,
\big[\,(\sin\alpha)^2\,-\,\Lambda_n(\alpha)\,\cos\alpha\,\big]\,\,.
\eeq
Using this result, the BPS condition can be written as:
\beq
\alpha'\,=\,{1\over r}\,\,
{\sin\alpha\,\Lambda_n(\alpha)\,\over (\sin\alpha)^2\,-\,\Lambda_n(\alpha)\,\cos\alpha}\,\,.
\label{BPS-eq}
\eeq
Several observations concerning (\ref{BPS-eq}) are in order.  First of all, notice that the solutions of (\ref{BPS-eq}) with constant $\alpha$ are those with $\alpha=0,\pi$ or with $\alpha$ being one of the zeroes of the function $\Lambda_n(\alpha)$. Thus, our flux tube embeddings are certainly a particular solution of the BPS equation (\ref{BPS-eq}). Moreover, one can verify by direct calculation that any solution $\alpha(r)$ of (\ref{BPS-eq})  also solves the second-order differential equation of motion (\ref{embedding-eom}). 
To prove this fact it is quite useful to demonstrate first that:
\beq
 \sqrt{1\,+\,r^2\,\big(\alpha'\big)^2}\,\,\Big|_{BPS}\,=\,
 {\sqrt{\big(\sin\alpha\big)^6\,+\,{\cal C}_n(\alpha)^2}\over 
  (\sin\alpha)^2\,-\,\Lambda_n\,\cos\alpha}\,\,.
  \label{sqrt-BPS}
 \eeq
Furthermore, by using (\ref{sqrt-BPS}) one can also prove that the electric field for a BPS configuration is given by:
\beq
F_{0r}\,\big|_{BPS}\,=\,{R^3\over 4k}\,\,\partial_r\,\big(\,
r\,\cos\alpha\,\big)\,\,.
\label{electric-BPS}
\eeq
Actually, one can easily demonstrate by using (\ref{F0r}) that the first-order BPS equation (\ref{BPS-eq}) for the embedding is equivalent to having an electric field given by (\ref{electric-BPS}). Moreover, although we have not verified it, it is likely that the first-order equation (\ref{BPS-eq}) and the electric field (\ref{electric-BPS}) could by derived from the kappa symmetry of the probe brane, in a calculation similar to the one done in \cite{Gomis:1999xs} for the  ${\cal N}=4$ baryon vertex. 
 
By using the fact that $d\Lambda_n(\alpha)/d\alpha=2\sin\alpha$, one can easily show that the BPS equation can be integrated exactly in the form:
\beq
r\,=\,C\,{\sqrt{-\Lambda_n(\alpha)}\over \sin\alpha}\,\,,
\eeq
where $C$ is a constant of integration. By redefining $C$ and using the expression of $\Lambda_n(\alpha)$ in (\ref{Lambda_n}), this solution can be written as:
\beq
r\,=\,C\,\,
{\sqrt{\cos\alpha\,+\,{2n\over N}\,-\,1}\over \sin\alpha}\,\,.
\label{spike}
\eeq
Notice that, in the solution (\ref{spike}), the coordinate $r\to\infty$ when $\alpha\to 0$, which corresponds to spike of the D6-brane probe which reaches the boundary of $AdS_4$. Moreover, the angle $\alpha$ takes values in the interval $0\le \alpha\le\alpha_n$, where the $\alpha_n$ are the angles defined in (\ref{minimum_condition}) (when $0<n<N$ the brane reaches the origin $r=0$ when $\alpha=\alpha_n$) .  By looking carefully at the energy of the spike one can check that it corresponds to $n$ fundamental strings reaching the boundary of the Anti-de-Sitter space.

In the particular case $n=N$ the maximum value of $\alpha$  is $\alpha=\pi$ (\ie\ $0\le\alpha\le \pi$ in this case) and the D6-brane wraps completely the ${\mathbb C}\,{\mathbb P}^3$. This configuration corresponds to the baryon vertex of the ABJM model. In this $n=N$ case the function $r(\alpha)$ can be simply written as:
\beq
r(\alpha)\,=\,{r_0\over \sin\big({\alpha\over 2}\big)}\,\,,
\qquad\qquad
0\le\alpha\le \pi\,\,,
\eeq
where $r_0=r(\alpha=\pi)$.

\renewcommand{\theequation}{\rm{B}.\arabic{equation}}
\setcounter{equation}{0}
\medskip

\section{Fluctuations of the probe D6-branes}
\label{fluctuations}
In this appendix we analyze the small perturbations around the flux tube and dimer configurations    at zero-temperature. The goal is to find the second order lagrangian for these fluctuations that was the starting point of sections \ref{Fluct} and \ref{dimer-fluct}. We start by studying the fluctuations of the impurity configuration of section \ref{flux-tubes}.

\subsection{Impurity fluctuations}
Let us perturb the configuration of the D6-brane probe as in (\ref{perturbation}) and let us expand the DBI+WZ action to second order  in the perturbations $\xi$, $f$ and $\chi$. We begin by obtaining the different components of the induced metric $g_{mn}$ at second order in the fluctuations. We write:
\beq
g\,=\,\bar g\,+\,\hat g\,\,,
\eeq
where $\bar g$ is the zero-order induced metric written in (\ref{induced-metric-unperturbed}). When any of the indices $m$ and $n$ is outside the $\tilde T^{1,1}$, the  metric perturbation $\hat g_{mn}$ takes the form:
\beq
\hat g_{mn}\,=\,{R^3\over 4k}\,\,\Big[\,
\partial_m\xi\,\partial_n\xi\,+\,r^2\,\partial_m\chi\,\partial_n\chi\,\,\Big]\,\,.
\eeq
On the other hand, if both indices belong to the $\tilde T^{1,1}$, one has to expand the 
$\sin{\alpha\over 2}$ and $\cos{\alpha\over 2}$ factors appearing in the internal metric up to second order in $\xi$ and one gets:
\beq
\hat g_{ij}\,=\,{R^3\over 4k}\,\,\Big[\,
\partial_i\xi\,\partial_j\xi\,+\,r^2\,\partial_i\chi\,\partial_j\chi\,+\,
\xi\,\hat g^{(1)}_{ij}\,+\,\xi^2\,\hat g^{(2)}_{ij}\,\,\Big]\,\,,
\eeq
where $g^{(1)}$ and $g^{(2)}$ are given by:
\bear
&&\hat g^{(1)}_{ij}\,d\gamma^{i}\,d\gamma^{j}\,=\,{1\over 2}\,\,\Big[
-\sin\alpha_n\,\big(\,d\theta_1^2\,+\,\sin^2\theta_1\,d\varphi_1^2\,\big)\,+\,
\sin\alpha_n\,\big(\,d\theta_2^2\,+\,\sin^2\theta_2\,d\varphi_2^2\,\big)\,+\,
\rc\rc
&&\qquad\qquad\qquad\qquad\qquad\qquad
+\,\sin\alpha_n\,\cos\alpha_n\,
\big(\,d\chi+\cos\theta_1 \,d\varphi_1\,+\,\cos\theta_2 \,d\varphi_2\,\big)^2\,\Big]
\,\,,\rc\rc
&&\hat g^{(2)}_{ij}\,d\gamma^{i}\,d\gamma^{j}\,=\,{1\over 4}\,\,\Big[
-\cos\alpha_n\,\big(\,d\theta_1^2\,+\,\sin^2\theta_1\,d\varphi_1^2\,\big)\,+\,
\cos\alpha_n\,\big(\,d\theta_2^2\,+\,\sin^2\theta_2\,d\varphi_2^2\,\big)\,+\,
\rc\rc
&&\qquad\qquad\qquad\qquad
+\,(\sin^2\alpha_n\,-\cos^2\alpha_n)\,
\big(\,d\chi+\cos\theta_1 \,d\varphi_1\,+\,\cos\theta_2 \,d\varphi_2\,\big)^2
\,\Big]\,\,.
\label{hat-g}
\eear
In order to expand the DBI D6-brane action (\ref{DBI-D6}), we notice that
the Born-Infeld  determinant can be written as:
\beq
\sqrt{-\det(g+{\cal F})}\,=\,\sqrt{-\det \big(\,\bar g \,+\,{
\bar f}\,\big)}\,
\sqrt{\det(1+X)}\,\,,
\label{detX}
\eeq
where the matrix $X$ is given by:
\beq
X\,\equiv\,\big(\,\bar g\,+\,{\bar f}\,\big)^{-1}\,\,
\big(\,\hat g\,+\,{f}\,\big)\,\,.
\eeq
To evaluate  the right-hand side of eq. (\ref{detX}), we shall use 
the expansion:
\beq
\sqrt{\det(1+X)}\,=\,1\,+\,{1\over 2}\,\tr X\,-\,{1\over 4}\,\tr X^2\,+\,
{1\over 8}\,\big(\tr X\big)^2\,+\,o(X^3)\,\,.
\label{expansion}
\eeq
Moreover, in the inverse  matrix
$\big(\,\bar g\,+\,{ \bar f}\,\big)^{-1}$ we will separate the symmetric and antisymmetric parts:
\beq
\big(\,\bar g\,+\,{\bar f}\,\big)^{-1}\,=\,{\cal G}^{-1}\,+\,
{\cal J}\,\,,
\label{openmetric}
\eeq
where ${\cal J}$ is the antisymmetric component. The symmetric matrix ${\cal G}$ is the so-called open string metric and is the one that naturally shows up in  the fluctuations of the worldvolume when gauge fields are turned on.

The matrix $\bar g\,+\,{\bar f}$ has a block structure. By computing  the inverse in the $0r$ sector (\ie\ in the $AdS_2$ part of the geometry), one gets:
\beq
\big(\,\bar g\,+\,{\bar f}\,\big)^{-1}_{|_{0r}}\,=\,{4k\over R^3\,\sin^2\alpha_n}\,\,
\left(  \begin{array}{cc}-r^{-2}
&\cos\alpha_n\\\\
-\cos\alpha_n&r^2
\end{array}\right)\,\,.
\label{bar-g+f}
\eeq
It follows from (\ref{bar-g+f}) that the non-vanishing elements of ${\cal J}$  are:
\beq
{\cal J}^{0r}\,=\,-{\cal J}^{r0}\,=\,{4k\over R^3}\,{\cot\alpha_n\over \sin\alpha_n}\,\,.
\eeq
In terms of the filling fraction $\nu=n/N$, the above expression of ${\cal J}^{0r}$ can be written as:
\beq
{\cal J}^{0r}\,=\,-{\cal J}^{r0}\,=\,{k\over R^3}\,
{N(N-2n)\over n(N-n)}\,=\,{k\over R^3}\,{1-2\nu\over \nu(1-\nu)}\,\,.
\eeq
Moreover, the non-vanishing elements of the inverse open string metric are:
\beq
{\cal G}^{00}\,=\,-{4k\over R^3}\,{1\over r^2\sin^2\alpha_n}\,\,,\qquad
{\cal G}^{rr}\,=\,{4k\over R^3}\,{r^2\over \sin^2\alpha_n}\,\,,\qquad
{\cal G}^{ij}\,=\,{4k\over R^3}\,\tilde g^{ij}\,\,,
\label{inverse-open-metric}
\eeq
and, thus, the open string metric takes the form:
\beq
{\cal G}_{mn}\,d\zeta^{m}\,d\zeta^{n}\,=\,{R^3\over 4k}\,\,
\big[\,\sin^2\alpha_n\,ds^2_{AdS_2}\,+\,d s^2_{\tilde T^{1,1}}\,\big]\,\,,
\label{open-metric}
\eeq
where $ds^2_{AdS_2}=-r^2\,dt^2\,+\,dr^2/r^2$ and $d s^2_{\tilde T^{1,1}}$ has been written in (\ref{squashedT11}) and (\ref{squashedT11-n}). 

Let us now compute the different terms on the right-hand side of (\ref{expansion}). The matrix elements that contribute to $\tr(X)$ are:
\bear
&&X^{0}_{\,\,\,\, 0}\,=\,-{4k\over R^3}\,
{\cot\alpha_n\over \sin\alpha_n}\,\,f_{0r}\,-\,
{1\over r^2 \sin^2\alpha_n}\,\,\big(\partial_0\xi\big)^2
\,-\,{1\over  \sin^2\alpha_n}\,\,\big(\partial_0\chi\big)^2
\,+\,
{\cot\alpha_n\over \sin\alpha_n}\,\partial_0\xi\,\partial_r\xi
,\,\,\rc\rc
&&X^{r}_{\,\,\,\, r}\,=\,-{4k\over R^3}\,
{\cot\alpha_n\over \sin\alpha_n}\,\,f_{0r}\,+\,
{r^{2}\over \sin^2\alpha_n}\,\,\big(\partial_r\xi\big)^2\,+\,
{r^{4}\over \sin^2\alpha_n}\,\,\big(\partial_r\chi\big)^2
\,-\,{\cot\alpha_n\over \sin\alpha_n}\,\partial_0\xi\,\partial_r\xi
\,\,,\rc\rc
&&X^{i}_{\,\,\,\, j}\,=\,\xi\,\big(\,{\cal M}^{(1)}\,\big)^i_{\,\,j}\,+\,
{4k\over R^3}\,\tilde g^{ik}\,f_{kj}+
\xi^2\,\big(\,{\cal M}^{(2)}\,\big)^i_{\,\,j}\,+\,
\tilde g^{ik} \partial_k\xi\, \partial_j\xi\,+\,
r^2\,\tilde g^{ik} \partial_k\chi\, \partial_j\chi\,\,,\qquad\qquad
\eear
where $i, j, k$ are indices along the internal directions and the matrices ${\cal M}^{(a)}$ are defined as:
\beq
\big(\,{\cal M}^{(a)}\,\big)^i_{\,\,j}\,\equiv\,\tilde g^{ik}\,\,\hat g^{(a)}_{kj}\,\,,
\qquad \qquad (a=1,2)\,\,,
\eeq
with $\tilde g$ being the $\tilde T^{1,1}$ metric (\ref{squashedT11}) and 
$\hat g^{(a)}$ are the metrics written in (\ref{hat-g}). One can verify that 
the matrices ${\cal M}^{(a)}$ are, in fact, diagonal. Actually, ordering the directions of the internal manifold as:
\beq
\gamma^i\,=\,(\theta_1,\theta_2,\varphi_1,\varphi_2,\chi)\,\,,
\eeq
it is easy to check that ${\cal M}^{(1)}$ and ${\cal M}^{(2)}$ are just:
\bear
&& {\cal M}^{(1)}\,=\,{\rm diag}\,\Big(-\tan\bigg({\alpha_n\over 2}\bigg)\,,\,
\cot\bigg({\alpha_n\over 2}\bigg)\,,\,
-\tan\bigg({\alpha_n\over 2}\bigg)\,,\,
\cot\bigg({\alpha_n\over 2}\bigg)\,,\,2\cot\alpha_n\Big)\,\,,\qquad\rc\rc
&& {\cal M}^{(2)}\,=\,{\rm diag}\,\Big(
-{\cos\alpha_n\over 4\cos^2({\alpha_n\over 2})}\,,\,
{\cos\alpha_n\over 4\sin^2({\alpha_n\over 2})}\,,\,
-{\cos\alpha_n\over 4\cos^2({\alpha_n\over 2})}\,,\,
{\cos\alpha_n\over 4\sin^2({\alpha_n\over 2})}\,,\,
\cot^2\alpha_n-1\Big)\,\,.\qquad\qquad
\eear
It follows that the traces of the ${\cal M}$'s are:
\beq
\tr\,\big( {\cal M}^{(1)}\big)\,=\,6\cot\alpha_n\,\,,\qquad\qquad
\tr\,\big( {\cal M}^{(2)}\big)\,=\,3\cot^2\,\alpha_n\,-\,1\,\,.
\eeq
From these expressions it is now straightforward to compute the trace of $X$, with the result:
\bear
&&\tr (X)\,=\,-{8k\over R^3}\,{\cot\alpha_n\over \sin\alpha_n}\,f_{0r}\,+\,
6\,\cot\alpha_n\,\xi\,+(3\cot^2\alpha_n\,-\,1)\,\xi^2\,+\,\rc\rc
&&\qquad\qquad\qquad\qquad
+\,{R^3\over 4k}\,{\cal G}^{mn}\,\partial_m\,\xi\,\partial_n\,\xi\,+\,
{R^3\over 4k}\,r^2\,{\cal G}^{mn}\,\partial_m\,\chi\,\partial_n\,\chi\,\,.
\eear
In order to calculate the trace of $X^2$, one needs to compute the non-diagonal elements of $X$. To the order we are working, it is enough to calculate them at first order. One gets:
\bear
&&X^{0}_{\,\,\,\, r}\,=\,-{4k\over R^3}\,{1\over r^2\sin^2\alpha_n}\,\,f_{0r}\,\,,
\qquad\qquad\qquad\qquad\qquad\,\,\,\,\,\,\,
X^{r}_{\,\,\,\, 0}\,=-{4k\over R^3}\,{r^2\over \sin^2\alpha_n}\,\,f_{0r}\,\,,\rc\rc
&&X^{0}_{\,\,\,\, i}\,=\,-{4k\over R^3}\,{1\over r^2\sin^2\alpha_n}\,\,f_{0i}\,+\,
{4k\over R^3}\,{\cot\alpha_n\over \sin \alpha_n}\,f_{ri}\,\,,
\qquad\qquad
X^{i}_{\,\,\,\, 0}\,=\,{4k\over R^3}\,\tilde g^{ij}\,f_{j0}\,\,,
\qquad\qquad\qquad
\rc\rc
&&X^{r}_{\,\,\,\, i}\,=\,{4k\over R^3}\,{r^2\over \sin^2\alpha_n}\,\,f_{ri}\,-\,
{4k\over R^3}\,{\cot\alpha_n\over \sin \alpha_n}\,f_{0i}\,\,,
\qquad\qquad\qquad
X^{i}_{\,\,\,\, r}\,=\,{4k\over R^3}\,\tilde g^{ij}\,f_{jr}\,\,.
\eear
Moreover, by using that:
\beq
\tr\,\big( {\cal M}^{(1)}\, {\cal M}^{(1)}\big)\,=\,4\,(3\cot^2\alpha_n\,+\,1\,)\,\,,
\eeq
one gets:
\bear
&&\tr (X^2)\,=\,{32 k^2\over R^6}\,\,{\cos^2 \alpha_n\,+\,1\over \sin^4 \alpha_n}
\,f_{0r}^2
\,+\, 4(3\cot^2\alpha_n\,+\,1\,)\,\xi^2\,+\,\rc\rc
&&\qquad\qquad\qquad
\,+\,{16 k^2\over R^6}\,\Big(\,{2\over r^2\,\sin^2 \alpha_n}\,f_{0i}^2\,-\,
{2 r^2\over \,\sin^2 \alpha_n}\,f_{ri}^2\,-\,f_{kj}^2\,\Big)\,\,.
\eear
With these results one can now readily compute $\sqrt{\det(1+X)}$, namely:
\bear
&&\sqrt{\det(1+X)}\,=\,
1\,-\,{4k\over R^3}\,{\cot\alpha_n\over \sin\alpha_n}\,f_{0r}\,+\,3\,\cot\alpha_n\,\xi\,+\,
{R^3\over 4k}\,{1\over 2}\,{\cal G}^{mn}\,\partial_m\,\xi\,\partial_n\,\xi\,+\,\rc\rc
&&\qquad\qquad\qquad\qquad
+\,{R^3\over 4k}\,{1\over 2}\,r^2\,{\cal G}^{mn}\,\partial_m\,\chi\,\partial_n\,\chi\,+\,
{1\over 4}\,{\cal G}^{mp}\,{\cal G}^{nq}\,f_{pq}\,f_{mn}\,+\,\rc\rc
&&\qquad\qquad\qquad\qquad
+\,{3\over 2}\,(2\cot^2\alpha_n\,-\,1\,)\,\xi^2\,-\,
{12k\over R^3}\,{\cot^2\alpha_n\over \sin\alpha_n}\,\xi\,f_{0r}\,\,.
\label{det(1+X)}
\eear
Let us now use this result to compute the DBI action for the fluctuations. First,  we notice that:
\beq
e^{-\phi}\,\sqrt{-\det (\bar g+\bar f)}\,=\,{R^9\over 2^7\,k^2}\,\,\sin\alpha_n\,
\sqrt{\tilde g}\,\,,
\eeq
where $\tilde g$ is the determinant of the metric of the squashed $T^{1,1}$ which, according to (\ref{squashedT11}),  is given by:
\beq
\sqrt{\tilde g}\,=\,{\sin^3\alpha_n\over 8}\,\,\sin\theta_1\,\sin\theta_2\,\,.
\eeq
Therefore, if we represent the DBI action as:
\beq
S_{DBI}\,=\,\int {\cal L}_{DBI}\,\,d^7\zeta\,\,,
\eeq
the DBI lagrangian density is given by:
\beq
{\cal L}_{DBI}\,=\,-T_{6}\,\,
{R^9\over 2^7\,k^2}\,\,\sin\alpha_n\,
\sqrt{\tilde g}\,\,\sqrt{\det(1+X)}\,\,,
\label{DBI-lagrangian-expanded}
\eeq
where $\sqrt{\det(1+X)}$ has been written in terms of the fluctuations in (\ref{det(1+X)}). 

The WZ lagrangian density has been written in (\ref{L-WZ}). Let us write in this equation
$F_{0r}={R^3\over 4k}\,\cos\alpha_n\,+\,f_{0r}$ and let us expand the function $C(\alpha)$ as:
\beq
C(\alpha)\,=\,C(\alpha_n)\,-\,3\,(\sin\alpha_n)^3\,\xi\,-{9\over 2}\,
(\sin\alpha_n)^2\,\cos\alpha_n\,\xi^2\,\,.
\eeq
Then, one gets:
\bear
&&{\cal L}_{WZ}\,=\,-T_{6}\,\,{R^6\over 2^8\,k}\,\,\sin\theta_1\,\sin\theta_2\,
\Big[\,{R_3\over 4k}\,C(\alpha_n)\,\cos\alpha_n\,+\,C(\alpha_n)\,f_{0r}\,-\,\rc\rc
&&\qquad\qquad
-{3R^3\over 4k}\,\sin^3\alpha_n\,\cos\alpha_n\,\xi\,-\,
{9R^3\over 8k}\,\sin^2\alpha_n\,\cos^2\alpha_n\,\xi^2\,-\,3 
\sin^3\alpha_n\,\xi\,f_{0r}\,\Big]\,\,.
\qquad
\label{WZ-Lagrangian-expanded}
\eear
The total lagrangian density for the fluctuations is just the sum of (\ref{DBI-lagrangian-expanded}) and (\ref{WZ-Lagrangian-expanded}). After neglecting terms that are constant or total derivatives, one can easily prove that the total lagrangian density is given by (\ref{lagrangian-fluctuations}).

\subsection{Dimer fluctuations}

Let us now study the fluctuations around the connected configuration given by the ansatz 
(\ref{dimer-ansatz}) in the zero-temperature  ABJM background of section \ref{ABJM}. First of all, we notice that the function $x(r)$ is given by:
\beq
{dx\over dr}\,=\,\pm\,{r_{*}^2\over r^2 \sqrt{r^4-r_{*}^4}}\,\,,
\label{x-prime-SUSY}
\eeq
with $r_*$ being the minimum of the coordinate $r$ reached by the brane.  Eq. (\ref{x-prime-SUSY}) is immediately obtained from (\ref{xprime}) by taking $f(r)=1$ and by redefining the constant $\hat\Lambda$ as $\hat\Lambda^2=r_*^4$ (see (\ref{turning})). 
The length $L$  of the dimer is now:
\beq
L\,=\,2\,r_{*}^2\,\int_{r_{*}}^{\infty}\,{dr\over r^2\sqrt{r^4-r_{*}^4}}\,\,.
\label{L-dimer-SUSY}
\eeq
By performing explicitly the integral on the right-hand-side of (\ref{L-dimer-SUSY}), we get:
\beq
L\,=\,{2\sqrt{2}\,\pi^{{3\over 2}}\over \big[\,\Gamma\big({1\over 4}\big)\,\big]^{2}}\,\,
{1\over r_*}\,\,.
\label{L-dimer-SUSY-explicit}
\eeq
The induced metric in this case becomes:
\beq
d\bar s^2\,=\,{R^3\over 4k}\,\,\Big[\,-r^2\,dt^2\,+\,{h(r)\over r^2}\,dr^2\,+\,
ds^2_{\tilde T^{1,1}}\,\Big]\,\,,
\eeq
where $h(r)$ is the function:
\beq
h(r)\,\equiv \,{1\over 1-{r_{*}^4\over r^4}}\,\,.
\eeq
Notice that in this case the metric is of the form $AdS_2\times \tilde T^{1,1}$ only asymptotically  in the UV. The worldvolume gauge field strength for this hanging configuration depends on the holographic variable $r$. Indeed, from (\ref{electric-field-dimer}) one easily gets:
\beq
\bar f_{0r}\,=\,{R^3\over 4k}\,\,\sqrt{h(r)}\,\cos\alpha_n\,\,.
\label{electric-field-dimer-SUSY}
\eeq
(Compare this result  with (\ref{f-unperturbed})). Proceeding as in the case of the unconnected configuration, it is straightforward to obtain the open string metric ${\cal G}_{mn}$ for this case. One gets:
\beq
{\cal G}_{mn}\,d\zeta^{m}\,d\zeta^{n}\,=\,{R^3\over 4k}\,\,
\Big[\,\sin^2\alpha_n\,
\big(-r^2\,dt^2\,+\,{h(r)\over r^2}\,dr^2\,\big)
\,+\,d s^2_{\tilde T^{1,1}}\,\Big]\,\,.
\label{open-metric-hanging}
\eeq

The lagrangian density for a generic fluctuation of the dimer is highly coupled and, for this reason, we will not try to study here the dimer fluctuations in full generality. In this paper we will restrict ourselves to a set of fluctuation modes which are decoupled from the others. In these modes the only coordinate that fluctuates is the cartesian coordinate transverse to the dimer direction in the Minkowski space (\ie\ $y=y_0+\chi$, with $y_0$ constant). It is straightforward to expand the DBI+WZ action of the probe and to obtain the lagrangian density to second order in $\chi$. One gets:
\beq
{\cal L}\,=\,-T_{6}\,\,{R^9\over 2^7\,k^2}\,\sin\alpha_n\,\sqrt{\tilde g}\,
\Big[\,{R^3\over 4k}\,{r^2\,\over 2}\,\sqrt{h}\,\,{\cal G}^{mn}\,\partial_m\,\chi\,\partial_n\,\chi\,\,\Big]\,\,.
\label{dimer-fluct-lagrangian}
\eeq
By comparing this expression with the first term in (\ref{lagrangian-fluctuations}) we notice that in (\ref{dimer-fluct-lagrangian}) there is an extra factor of $\sqrt{h}$, and the difference between the lagrangians is similar to the one  between the open string metrics (\ref{open-metric-hanging}) and (\ref{open-metric}). 
The equation of motion derived from (\ref{dimer-fluct-lagrangian})  is:
\beq
\partial_m\,\Big[\,r^2\sqrt{\tilde g}\,\sqrt{h}\,\,{\cal G}^{mn}\,\partial_n\,\chi\,\Big]\,=\,0\,\,.
\label{chi-eq-hanging}
\eeq
This equation is studied in detail in section \ref{dimer-fluct}.

\vskip 1cm
\renewcommand{\theequation}{\rm{C}.\arabic{equation}}
\setcounter{equation}{0}
\medskip

\section{Laplacian for the squashed $T^{1,1}$}
\label{harmonics}

Let us consider a five-dimensional metric of the type:
\beq
ds^2\,=\,A\,(\,d\theta_1^2\,+\,\sin^2\theta_1\,d\phi_1^2\,)\,+\,
B\,(\,d\theta_1^2\,+\,\sin^2\theta_1\,d\phi_1^2\,)\,+\,
C\,(\,d\chi+\cos\theta_1\,d\phi_1\,+\,\cos\theta_2\,d\phi_2\,)^2\,\,,
\label{T11-squashed}
\eeq
where $A$, $B$ and $C$ are constants. The laplacian operator for the metric (\ref{T11-squashed})  acts on a scalar function $H$ as:
\beq
\nabla^2\,H\,=\,{1\over \sqrt{g}}\partial_m\,\Big(\sqrt{g}\,g^{mn}\partial_n \,H\Big)\,\equiv\,
-\Delta_0\,H\,\,,
\eeq
where, for convenience, we have defined the operator $\Delta_0=-\nabla^2$ ($\Delta_0$ is the Hodge-de-Rham operator acting on zero-forms).  Let us  now define the following operators:
\bear
&&
\nabla^2_i\equiv {1\over \sin\theta_i}\,
\partial_{\theta_i}\big(\sin\theta_i\partial_{\theta_i}\,\big)\,+\,
\Big({1\over \sin\theta_i}\,\partial_{\phi_i}\,-\,\cot\theta_i\,\partial_{\chi}\,\Big)^2
\,\,,\qquad(i=1,2)\,\,,\rc\rc
&&\nabla^2_R\,=\,\partial^2_{\chi}
\eear
Then, the laplacian $\Delta_0$ can be expressed in terms of $\nabla^2_{1,2}$ and $\nabla^2_R$  as:
\beq
-\Delta_0\,=\,{1\over A}\,\nabla^2_1\,+\,{1\over B}\,\nabla^2_2\,+\,{1\over C}\,\nabla^2_R\,\,.
\eeq
The laplacian operator for a round three-sphere is given by:
\beq
\nabla^2_{{\mathbb  S}^3}\,=\,4\,
\Big[\, {1\over \sin\theta}\,
\partial_{\theta}\big(\sin\theta\partial_{\theta}\,\big)\,+\,
\Big({1\over \sin\theta}\,\partial_{\phi}\,-\,\cot\theta\,\partial_{\chi}\,\Big)^2\,+\,
\partial^2_{\chi}\,\Big]\,\,,
\eeq
and its eigenvalues are $-l(l+2)$ with $l\in \mathbb{Z}$. It follows  that the operators 
$\nabla^2_i$ are related to the laplacian $\nabla^2_{{\mathbb S}^3_i}$ of a three-sphere parametrized  by  $(\theta_i,\phi_i, \chi)$  by means of the relation:
\beq
\nabla^2_i\,=\,{1\over 4}\,\nabla^2_{{\mathbb S}^3_i}\,-\,\partial^2_{\chi}\,\,.
\eeq
The eigenfunctions of $\partial^2_{\chi}$ are of the form $e^{i{r\over 2}\chi}$ with 
$r\in \mathbb{Z}$ and the eigenvalues are of the form $-r^2/4$.  Then, the eigenvalues of $\nabla^2_i$ are:
\beq
-{l_i(l_i+2)\over 4}\,+\,{r^2\over 4}\,\,,\qquad
l_i,r\in \mathbb{Z}\,\,.
\eeq
Therefore, the eigenvalues of $\Delta_0$ are:
\beq
H_0(l_1,l_2, r)\,=\,
{l_1(l_1+2)\over 4\, A}\,+\,{l_2(l_2+2)\over 4\, B}\,-\,
\Big[\,{1\over A}\,+\,{1\over B}\,-\,{1\over C}\,\Big]\,{r^2\over 4}\,\,.
\label{H0-general-squashing}
\eeq
Let us apply this result to the case of the squashed $T^{1,1}$ metric written in (\ref{squashedT11}). In this case the values of $A$, $B$ and $C$ are:
\beq
A\,=\,\cos^2{\alpha_n\over 2}\,\,,\qquad
B\,=\,\sin^2{\alpha_n\over 2}\,\,,\qquad
C\,=\,\sin^2{\alpha_n\over 2}\,\cos^2{\alpha_n\over 2}\,\,.
\eeq
In terms of $n$ and $N$ the actual values of the coefficients $A$, $B$ and $C$ are (see (\ref{squashedT11-n})):
\beq
A\,=\,{N-n\over N}\,\,,
\qquad\qquad
B\,=\,{n\over N}\,\,,
\qquad\qquad
C\,=\,{n(N-n)\over N^2}\,\,.
\eeq
These values satisfy:
\beq
{1\over A}+{1\over B}\,=\,{1\over C}\,\,,
\eeq
and, therefore,  the eigenvalues of the laplacian are independent of $r$ and given by:
\beq
H_0\,=\,{N\over 4}\,\Big[\,{l_1(l_1+2)\over N-n}\,+\,{l_2(l_2+2)\over n}\,\Big]\,\,.
\label{H0-particular-squashing}
\eeq

\vskip 1cm
\renewcommand{\theequation}{\rm{D}.\arabic{equation}}
\setcounter{equation}{0}
\medskip

\section{Harmonic calculus}
\label{harmonic}

In this appendix we develop the harmonic calculus for a squashed $T^{1,1}$ space. We will follow closely the algebraic approach of refs. \cite{Ceresole:1999ht,Ceresole:1999zs}, in which the $T^{1,1}$ space is realized as  a coset of the type $SU(2)\times SU(2)/U_H(1)$. We denote by $T_1$, $T_2$ and $T_3$ the generators of the first $SU(2)$ and by  $\hat T_1$, $\hat T_2$ and $\hat T_3$ the generators of the second $SU(2)$. The $U_H(1)$ is generated by:
\beq
T_H\,=\,T_3\,+\,\hat T_3\,\,.
\eeq
Moreover, let us introduce $T_5$ as:
\beq
T_5\,=\,T_3\,-\,\hat T_3\,\,.
\eeq
In the following we will adopt the notations used in refs. \cite{Ceresole:1999ht,Ceresole:1999zs}. The indices $a,b,c\cdots=1,\cdots,5$ will be used to denote a coset direction, $i,j, k\cdots=1,2$ will refer to the directions along $T_{1,2}$, while the indices $r,s,t\cdots=3,4$  will denote the directions along $\hat T_{1,2}$. The internal indices have a negative definite metric. We will also work with the combinations $T_{\pm}$ and $\hat T_{\pm}$, defined as:
\beq
T_{\pm}\,=\,T_1\,\pm i\,T_2\,\,,
\qquad\qquad
\hat T_{\pm}\,=\,\hat T_1\,\pm i\,\hat T_2\,\,.
\eeq
In order to describe the coset geometry, 
we will use a basis of rescaled vielbeins $V^{a}\,=\,(V^i\,,\,V^s\,,\,V^5)$ with the rescaling factors  $r(a)$ being given by:
\beq
r(i)\,=\,a\,\,,\qquad\qquad
r(s)\,=\,b\,\,,\qquad\qquad
r(5)\,=\,c\,\,.
\eeq
The vielbeins $V^{a}$ satisfy the torsion-free condition:
\beq
dV^{a}\,=\,{\cal B}^{ab}\,V_{b}\,\,,
\eeq
where ${\cal B}^{ab}$ is the Riemann connection one-form (${\cal B}^{ab}=-{\cal B}^{ba}$ for $a,b=i,s,5$). The $SO(5)$ covariant derivative is defined as:
\beq
{\cal D}\,=\,d+\,{\cal B}^{ab}\,T_{ab}\,\,,
\eeq
with $T_{ab}$ being the $SO(5)$ generators. As explained in detail in 
\cite{Ceresole:1999ht,Ceresole:1999zs}, the covariant derivative ${\cal D}$ acts algebraically on the coset representatives, which can be taken as the basic harmonics. 
Indeed, let $M^{ab}$ denote the part of ${\cal B}^{ab}$ orthogonal to the so-called $H$-connection. Then, the covariant derivative acting on the coset space can be written as:
\beq
{\cal D}\,=\,-r(a)\,V^{a}\,T_{a}\,+\,M^{ab}\,T_{ab}\,\,.
\eeq
In order to write explicitly the form of ${\cal D}$, let us notice that any $SO(5)$ representation can be branched with respect to its $U(1)_H$ subgroup. In particular, the vielbein $V^{a}$, which transforms in the $SO(5)$ vector representation, can be decomposed into five one-dimensional fragments $V^i=(V^1,V^2)$, $V^s=(V^3,V^4)$ and $V^5$ with $U(1)_H$ charges given respectively by $(1,-1)$, $(1,-1)$ and $0$, Then, the components of the covariant derivative ${\cal D}$  acting on harmonics are:
\bear
&&{\cal D}_i\,=\,-a\,T_i\,-\,{a^2\over 2c}\,\epsilon_i^{\,\,j}\,\,T_{5j}\,\,,\rc\rc
&&{\cal D}_s\,=\,-b\,T_s\,+\,{b^2\over 2c}\,\epsilon_s^{\,\,t}\,\,T_{5t}\,\,,\rc\rc
&&{\cal D}_5\,=\,-c\,T_5\,-\,2\,\Big(\,c-{a^2\over 4c}\,\Big)\,T_{12}\,+\,
2\,\Big(\,c-{b^2\over 4c}\,\Big)\,T_{34}\,\,.
\eear

The second-order Laplace-Beltrami operator is just ${\cal D}_a\,{\cal D}^a$. Let us first compute its eigenvalues acting on scalar harmonics. In this case the $SO(5)$ representation is trivial and ${\cal D}_a\,{\cal D}^a$ reduces to $\Delta_0$, where 
$\Delta_0$ is given by:
\beq
\Delta_0\,\equiv\,-a^2\,T_i\,T_i\,-\,b^2\,T_s\,T_s\,-\,c^2\,T_5\,T_5\,\,.
\eeq
In order to find the eigenvalues of $\Delta_0$, let us consider the action of the $T^{a}$ generators on the spherical harmonics. The latter are denoted by $Y_{q}^{(j,l,r)}$, where $j$ and $l$ are the spin quantum numbers of the two $SU(2)$ ($j,l\in {\mathbb Z}/2$), $r$ is the charge under the $U(1)_R$ generated by $T_5$ and $q$ is the charge under 
$U(1)_H$.  The generators $T_{\pm}$, $\hat T_{\pm}$ and $T_{5}$ act on the $Y$'s as follows:
\bear
&&T_{\pm}\,Y_{q}^{(j,l,r)}\,=\,-i\,
\Big(j\mp{q+r\over 2}\Big)\,Y_{q\pm 1}^{(j,l,r\pm 1)}\,\,,\rc\rc
&&\hat T_{\pm}\,Y_{q}^{(j,l,r)}\,=\,-i\,
\Big(l\mp {q-r\over 2}\Big)\,Y_{q\pm 1}^{(j,l,r\mp 1)}\,\,,\rc\rc
&&T_{5}\,Y_{q}^{(j,l,r)}\,=\,i\,r\,Y_{q}^{(j,l,r)}\,\,.
\eear
From this equation one gets:
\bear
&&T_i\,T_i\,Y_{q}^{(j,l,r)}\,=\,-\,\Big[\,j(j+1)\,-\,{(q+r)^2\over 4}\,\Big]\,
\,Y_{q}^{(j,l,r)}\,\,,\rc\rc
&&T_s\,T_s\,Y_{q}^{(j,l,r)}\,=\,-\,\Big[\,l(l+1)\,-\,{(q-r)^2\over 4}\,\Big]\,
\,Y_{q}^{(j,l,r)}\,\,,\rc\rc
&&T_5\,T_5\,Y_{q}^{(j,l,r)}\,=\,-r^2\,Y_{q}^{(j,l,r)}\,\,.
\eear
Then, one can demonstrate easily that:
\beq
\Delta_0\,Y_{q}^{(j,l,r)}=
\Big[\,a^2\,j(j+1)+b^2\,l(l+1)+(4c^2-a^2-b^2)\,{r^2\over 4}-
(a^2+b^2)\,{q^2\over 4}-(a^2-b^2)\,{q r\over 2}\,\Big]\,
Y_{q}^{(j,l,r)}\,\,.
\eeq
The scalar harmonics correspond to taking $q=0$. In this case the operator
$\Delta_0$ acts as:
\beq
\Delta_0\,Y_{0}^{(j,l,r)}\,=\,H_0\,Y_{0}^{(j,l,r)}\,\,,
\eeq
where the eigenvalue $H_0$ is given by:
\beq
H_0\,\equiv\,a^2\,j(j+1)\,+\,b^2\,l(l+1)\,+\,{4c^2-a^2-b^2\over 4}\,\,r^2\,\,.
\eeq

Let us now consider the Hodge-de-Rham operator $\Delta_1$ which, in the notation of \cite{Ceresole:1999ht,Ceresole:1999zs}, acts on the vector harmonics $Y_{(a)}$ as:
\beq
\Delta_1\,Y_{(a)}\,=\,{\cal D}_b\,{\cal D}^b\,Y_{(a)}\,+\,2 R_{a}^{\,\,b}\,Y_{(b)}\,\,,
\eeq
where $R_{a}^{\,\,b}$ is the Ricci tensor. Actually, for the $SU(2)\times SU(2)/U(1)$ coset metrics,  the Ricci tensor  is block diagonal and its components are given by:
\beq
R^i_{\,\,j}\,=\,\Big(\,{a^2\over 2}\,-\,{a^4\over 16\, c^2}\,\Big)\,\delta^i_k\,\,,\qquad
R^s_{\,\,t}\,=\,\Big(\,{b^2\over 2}\,-\,{b^4\over 16\, c^2}\,\Big)\,\delta^s_t\,\,,\qquad
R^5_{\,\,5}\,=\,{a^2\,b^2\over 8c^2}\,\,.
\eeq
Generically, $\Delta_1$ acts non-diagonally on the vector harmonics $Y_{(a)}$. Let us represent its action in terms of a matrix $\hat {\cal M}$:
\beq
\Delta_1\,Y_{(a)}\,=\,\hat{\cal M}_{a}^{\,\,\,\,b}\,Y_{(b)}\,\,.
\eeq
We will represent the generators of $SO(5)$ in the fundamental representation by means of the matrices
 $(T_{ab})^{cd}\,=\,{1\over 2}\,(\delta_a^c\,\delta_b^d\,-\,\delta_a^d\,\delta_b^c)$.
Moreover,  in order to write the components of $\hat {\cal M}$ it is convenient to use  as in  \cite{Ceresole:1999ht,Ceresole:1999zs} a complex basis for the vector representation, by defining the components  
$(\pm)=1\,\pm i\,2$ and $(\hat \pm)=3\,\pm i\,4$.  In this basis the only   non-vanishing elements of the matrix $\hat {\cal M}$ are:
\bear
&&\hat {\cal M}_{\pm}^{\,\,\,\,\pm}\,=\,\Delta_0\,+\,c^2\,+\,{a^2\over 2}\,\mp
2ic\Big(c\,-\,{a^2\over 4c}\,T_5\,\Big)\,\,,
\qquad\qquad\qquad
\hat {\cal M}_{\pm}^{\,\,\,\,5}\,=\,\pm i\,{a^3\over 4c}\,T_{\mp}\,\,,\rc\rc
&&\hat {\cal M}_{\hat\pm}^{\,\,\,\,\hat\pm}\,=\,\Delta_0\,+\,c^2\,+\,{b^2\over 2}\,\pm
2ic\Big(c\,-\,{b^2\over 4c}\,T_5\,\Big)\,\,,
\qquad\qquad\qquad
\hat {\cal M}_{\hat\pm}^{\,\,\,\,5}\,=\,\mp i\,{b^3\over 4c}\,\hat T_{\mp}\,\,,\rc\rc
&&\hat {\cal M}_{5}^{\,\,\,\,\pm}\,=\,\pm i\,{a^3\over 2c}\,\hat T_{\pm}\,\,,
\qquad\qquad
\hat {\cal M}_{5}^{\,\,\,\,\hat\pm}\,=\,\mp i\,{b^3\over 2c}\,\hat T_{\pm}\,\,,
\qquad\qquad
\hat {\cal M}_{5}^{\,\,\,\,5}\,=\,\Delta_0\,+\,{(a^2+b^2)^2\over 8c^2}\,\,.
\qquad\qquad
\eear
The  elements of $\hat {\cal M}$ contain generators $T_a$ which act non-trivially on the harmonics $Y_{(a)}$. Let us represent this action in terms of the matrix 
$ {\cal M}$ as:
\beq
\hat {\cal M}_{a}^{\,\,\,\,b}\,Y_{(b)}\,=\,
 {\cal M}_{a}^{\,\,\,\,b}\,Y_{(b)}\,\,.
 \eeq
To compute the values of the different elements of $ {\cal M}$ we have to specify the particular harmonics which compose our vector $Y_{(a)}$. Let us follow again 
\cite{Ceresole:1999ht,Ceresole:1999zs} and define:
\beq
Y_{(\pm)}\,\equiv\,Y_{\pm 1}^{(j,l,r\pm 1)}\,\,,\qquad
Y_{(\hat \pm)}\,\equiv\,Y_{\pm 1}^{(j,l,r\mp 1)}\,\,,\qquad
Y_{(0)}\,\equiv\,Y_{0}^{(j,l,r)}\,\,.
\eeq
We organize the harmonics as the following vector:
\beq
Y_{(a)} =\left(  \begin{array}{c}Y_{(+)} \\ Y_{(-)} \\ Y_{(\hat +)} \\   Y_{(\hat -)} \\ Y_{(0)}
 \end{array}\right)\,\,.
\eeq
Then, then the non-zero  elements of ${\cal M}_{a}^{\,\,\,\,b}$ are:
\bear
&&{\cal M}_{\pm}^{\,\,\,\,\pm}\,=\,H_0\,\mp\,{a^2\over 2}\,r
\,\,,\qquad\qquad
{\cal M}_{\pm}^{\,\,\,\,5}\,=\,\mp \,{a^3\over 4c}\Big(j\mp{r\over 2}\Big)\,\,,
\rc\rc
&&{\cal M}_{\hat\pm}^{\,\,\,\,\hat\pm}\,=\,H_0\,\pm\,{b^2\over 2}\,r
\,\,,\qquad\qquad
{\cal M}_{\hat\pm}^{\,\,\,\,5}\,=\,\pm \,{b^3\over 4c}\Big(l\pm{r\over 2}\Big)\,\,,
\rc\rc
&& {\cal M}_{5}^{\,\,\,\,\pm}\,=\,\mp {a^3\over 2c}\Big(j\pm {r\over 2}+1\Big)
\,\,,\qquad\qquad
 {\cal M}_{5}^{\,\,\,\,\hat\pm}\,=\,\pm {b^3\over 2c}\Big(l\mp {r\over 2}+1\Big)
\,\,,\rc\rc
&&{\cal M}_{5}^{\,\,\,\,5}\,=\,H_0\,+\,{(a^2+b^2)^2\over 8c^2}\,\,.
\eear
The above equations generalize the ones in \cite{Ceresole:1999ht,Ceresole:1999zs}  for general squashing factors $a$, $b$ and $c$. One can check that the results of 
 \cite{Ceresole:1999ht,Ceresole:1999zs}  are recovered when $a^2=b^2=6$ and $c=-3/2$.

\subsection{The ABJM harmonics}

Let us particularize our results for the metric of the $\tilde T^{1,1}$ space. In order to find the rescaling factors  $a$, $b$ and $c$ in terms of the filling fraction $\nu$, we compare the eigenvalues of $\Delta_0$ obtained above with those obtained directly from the differential operator. It is easy to see that $a$, $b$ and $c$ should be taken as:
\beq
a^2\,=\,{1\over 1-\nu}\,\,,
\qquad\qquad
b^2\,=\,{1\over \nu}\,\,,
\qquad\qquad
c^2\,=\,{1\over 4\nu (1-\nu)}\,\,.
\eeq
Indeed, for these values $a^2+b^2\,=\,4c^2$ and the eigenvalues $H_0$ do not depend on the R-charge $r$. Actually,  in this case $H_0$ is given by:
\beq
H_0\,=\,{j(j+1)\over 1-\nu}\,+\,{l(l+1)\over \nu}\,\,.
\eeq
In order to compare this result with the eigenvalues obtained from the differential operator, we should relate $j$ and $l$ with $l_1$ and $l_2$. This relation is just 
$j=l_1/2$, $l=l_2/2$ and, therefore, $H_0$ becomes:
\beq
H_0\,=\,{1\over 4}\,{l_1(l_2+2)\over 1-\nu}\,+\,{1\over 4}\,{l_2(l_2+2)\over \nu}\,\,,
\eeq
which, indeed, corresponds to the values found in appendix \ref{harmonics}  by diagonalizing the Laplacian operator. 

Let us now turn to the diagonalization of the matrix ${\cal M}$ of the Hodge-de-Rham operator.  The eigenvalues $\lambda$ are the roots of the polynomial equation:
\beq
p(\lambda)\equiv \det \big({\cal M}-\lambda\,I\big)\,=\,0\,\,,
\eeq
where $I$ is the $5\times 5$ unit matrix. Notice that  $p(\lambda)$ is a polynomial of degree five and, therefore, this equation has five roots. We are interested in finding the eigenvalues corresponding to modes that satisfy the transversality condition
${\cal D}^a\,Y_{(a)}\,=\,0$. Since we are not imposing this condition in our diagonalization problem, we are also including longitudinal harmonics of the form ${\cal D }_{a}\,Y$. The eigenvalue $\lambda$ for this longitudinal harmonic should be the same as the eigenvalue of $\Delta_0$ acting on a scalar harmonic, \ie\ $\lambda=H_0$. It is easy to determine when $\lambda=H_0$ is a solution of the eigenvalue equation. Indeed, one can check that:
\beq
p(\lambda=H_0)\,=\,-{r^4\over 32}\,\,{(1-2\nu)^2\over \nu^3\,(1-\nu)^3}\,\,.
\eeq
It follows that only for $r=0$ or $\nu=1/2$ (half-filling) is $\lambda=H_0$ a solution of the secular equation $p(\lambda)=0$. In these cases it is easy to identify the transverse eigenmodes as those orthogonal to the longitudinal one. For this reason we will restrict our analysis to these two cases.

\subsubsection{Arbitrary filling}

In this case we have to take $r=0$. The eigenvalue $\lambda=H_0$ appears three times. Assuming that one of these modes is the longitudinal one, two transverse modes with $\lambda=H_0$ remain. Therefore,  we get two series of eigenvalues:
\bear
&&\lambda_1\,=\,H_0\,\,,\rc\rc
&&\lambda_2^{(\pm)}\,=\,
{1\over 4\nu(1-\nu)}\,+\,H_0\,\pm\,
\sqrt{{1\over 16\nu^2(1-\nu)^2}\,+\,{\nu\over 4(1-\nu)^2}\,l_1(l_1+2)\,+\,
{1-\nu\over 4 \nu^2}\,l_2(l_2+2)}\,\,,
\qquad\qquad
\label{lambda12}
\eear
with $\lambda_1$ being doubly degenerate. 

\subsubsection{Half filling}
In this case $\nu={1\over 2}$ and $r$ can be arbitrary and we get two branches of eigenvalues:
\bear
&&\tilde\lambda_1^{(\pm)}\,=\,{l_1(l_1+2)\over 2}+\,{l_2(l_2+2)\over 2}\,\pm r\,\,,\rc\rc
&&\tilde\lambda_2^{(\pm)}\,=\,
1\,+\,{l_1(l_1+2)\over 2}+\,{l_2(l_2+2)\over 2}\pm\,
\sqrt{1\,+\,{l_1(l_1+2)\over 2}+\,{l_2(l_2+2)\over 2}}\,\,.
\qquad\qquad
\label{tilde-lambda12}
\eear
Notice that the eigenvalues $ \tilde\lambda_2^{(\pm)}$ are independent of $r$ and they are identical to the $\lambda_2^{(\pm)}$ for the particular case $\nu={1\over 2}$.

\vskip 1cm
\renewcommand{\theequation}{\rm{E}.\arabic{equation}}
\setcounter{equation}{0}

\section{The Heun equation}
\medskip
\label{Heun}

The Heun equation is the second-order differential equation given by:
\beq
{d^2\chi\over dz^2}\,+\,\Big(\,{\gamma\over z}\,+\,{\delta\over z-1}\,+\,
{\epsilon\over z-d}\,\Big)\,{d\chi \over dz}\,+\,
{\alpha\beta z\,-\,q\over z(z-1)(z-d)}\,\chi\,=\,0\,\,,
\label{Heun-eq}
\eeq
with the condition:
\beq
\gamma+\delta+\epsilon\,=\,\alpha+\beta+1\,\,.
\label{Heun-condition}
\eeq
The Heun equation is the standard canonical form of a Fuchsian equation with four singularities (located at $z=0, 1, d, \infty$). Recall that the Fuchsian equation with three singularities is the hypergeometric differential equation. The exponents of the singularities (which are the roots of the indicial equations) at $z=0,1,d,\infty$ are, respectively, $(0,1-\gamma)$, $(0,1-\delta)$, $(0,1-\epsilon)$ and $(\alpha,\beta)$. The parameter $q$ in (\ref{Heun-eq}) is called the accessory parameter.
The solution of (\ref{Heun-eq}) which is analytic  at $z=0$ is called the local Heun function and is denoted by $Hl(d,q;\alpha,\beta,\gamma,\delta;z)$ (see \cite{Ronveaux} for a review).

In some cases, the solutions of the Heun equation can be related to other special functions. So, when $\gamma=\delta=\epsilon\,=\,1/2$ the Heun equation reduces to the Lam\'e equation.  Moreover, when $d=2$, $\epsilon=\gamma$ and $q=\alpha\beta$, the local Heun function is related to the hypergeometric function by means of the equation:
\beq
Hl(2,\alpha\beta;\alpha,\beta,\gamma,\alpha+\beta-2\gamma+1;z)\,=\,
F\Big(\,{\alpha\over 2},{\beta\over 2};\gamma; z(2-z)\,\Big)\,\,.
\label{heun-hyper}
\eeq
This equivalence with hypergeometric functions seems to apply in our fluctuation equations with $\bar E=0$.

By performing a change of variables the Heun equation can  be converted into a Schr\"odinger equation for the so-called $BC_1$ elliptic Inozemtsev model, which is a quantum integrable model (see, for example refs.  \cite{Smirnov,Takemura, Maier}). Let us review this mapping, which involves elliptic functions.  With this purpose, let us briefly recall some basic facts about these functions. We will use the conventions of \cite{NIST}, where more details can be found. 

Let $\omega_1$ and $\omega_2$ be nonzero complex numbers such that 
 ${\rm Im}\,(\omega_3/ \omega_1)\,>0$. The lattice ${\mathbb L}$ with generators 
 $2\omega_1$ and $2\omega_3$ is defined as:
\beq
{\mathbb L}\,=\,\{\,w\in {\mathbb C}/ w=2n\,\omega_1\,+\,2m\,\omega_3\,,\,n,m\in {\mathbb Z}\,\}\,\,.
\eeq
The lattice parameter $\tau$ is defined as:
\beq
\tau\,=\,{\omega_3\over \omega_1}\,\,, \qquad\qquad
{\rm Im}\,\tau>0\,\,.
\label{tau-omegas}
\eeq
For a given lattice ${\mathbb L}$ we define the Weierstrass elliptic function 
$\wp(x)\,=\,\wp(x|{\mathbb L})$ by means of the formula:
\beq
\wp(x)\,=\,{1\over x^2}\,+\,\sum_{w\in {\mathbb L}\setminus \{0\}}\,
\Big(\,{1\over (x-w)^2}\,-\,{1\over w^2}\,\Big)\,\,.
\label{wp}
\eeq
We will also define $\omega_2$ as 
\beq
\omega_2\,=\,-\omega_1-\omega_3\,\,,
\eeq
and the numbers $e_1$, $e_2$ and $e_3$ as:
\beq
e_i\,=\,\wp(\omega_i)\,\,,\qquad\qquad
(i=1,2,3)\,\,,
\eeq
which satisfy the condition $e_1+e_2+e_3=0$.
In order to relate these quantities to the Jacobian elliptic functions, let us define the modulus $k$ and the complementary modulus $k'$ as:
\beq
k^2\,=\,{e_2-e_3\over e_1-e_3}\,\,,\qquad\qquad
k'^{\,2}\,=\,{e_1-e_2\over e_1-e_3}\,\,.
\eeq
From this definition one can check that $k$ and $k'$ satisfy the relation $k^2+k'^{\,2}\,=\,1$. These moduli can be related to the lattice parameter $\tau$ by means of the relation:
\beq
\tau\,=\,i\,{K(k')\over K(k)}\,\,,
\label{tau-modulus}
\eeq
where $K(k)$ is the complete elliptic integral of the first kind, defined as:
\beq
K(k)\,\equiv\,\int_0^{{\pi\over 2}}\,\,{d\theta\over \sqrt{1-k^2\,\sin^2\theta}}\,\,.
\eeq
In terms of this function and the moduli $k$ and $k'$, the numbers $e_1$, $e_2$ and $e_3$ are given by:
\beq
e_1\,=\,{1+k'^{\,2}\over 3\, \omega_1^2}\,\,\big[\,K(k)\,\big]^2\,\,,\qquad
e_2\,=\,{k^{\,2}-k'^{\,2}\over 3\, \omega_1^2}\,\,\big[\,K(k)\,\big]^2\,\,,\qquad
e_3\,=\,-{1+k^{\,2}\over 3 \,\omega_1^2}\,\,\big[\,K(k)\,\big]^2\,\,.\qquad
\label{es-K}
\eeq
The Weierstrass function $\wp(x)$ satisfies the differential equation:
\beq
\wp'(x)^2\,=\,4\,\wp(x)^3\,-\,g_2\,\wp(x)\,-\,g_3\,\,,
\label{wp-dif-eq}
\eeq
where $g_2$ and $g_3$ are the so-called lattice invariants, which are related to the
$e_i$ as follows:
\beq
g_2\,=\,2(\,e_1^2\,+\,e_2^2\,+\,e_3^2\,)\,\,,\qquad\qquad
g_3\,=\,4\,e_1\,e_2\,e_3\,\,.
\eeq

To convert the Heun equation into the Schr\"odinger equation of the Inozemtsev model,
let us consider elliptic functions with complementary modulus $k'$ given by:
\beq
k'^{\,2}\,=\,{1\over d}\,\,,
\label{modulus-d}
\eeq
where $d$ is parameter of the equation (\ref{Heun-eq}). 
We change variables from the $z$ of eq. (\ref{Heun-eq}) to a new variable $x$, related to 
$z$ by means of the following relation:
\beq
z\,=\,{\wp(x)\,-\,e_1\over e_2-e_1}\,\,.
\label{z-x-general}
\eeq
In terms of $x$, the Heun equation becomes a Schr\"odinger equation of the type:
\beq
H\,\Psi(x)\,=\,{\cal E}\,\Psi(x)\,\,,
\label{I-Sch}
\eeq
where the Hamiltonian $H$ is of the form:
\beq
H\,=\,-{d^2\over dx^2}\,+\,\sum_{i=0}^{3}\,l_i\,(l_i+1)\,\,
\wp (x+\omega_i)\,\,.
\label{I-hamiltonian}
\eeq
The coupling constants $l_i$ in (\ref{I-hamiltonian})  are related to the parameters of the Heun equation by means of the relations:
\beq
l_0\,=\,\beta\,-\,\alpha\,-{1\over 2}\,\,,\qquad
l_1\,=\,-\gamma\,+\,{1\over 2}\,\,,\qquad
l_2\,=\,-\delta\,+\,{1\over 2}\,\,,\qquad
l_3\,=\,-\epsilon\,+\,{1\over 2}\,\,,\qquad
\eeq
while the new function $\Psi(x)$ in (\ref{I-Sch}) and the old function $\chi(z)$ in (\ref{Heun-eq}) are related as:
\beq
\Psi(x)\,=\,z^{-{l_1\over 2}}\,\,(z-1)^{-{l_2\over 2}}\,\,(z-d)^{-{l_3\over 2}}\,\,\chi(z)\,\,.
\eeq
The ``energy" ${\cal E}$ of the equivalent Schr\"odinger problem is just:
\beq
{{\cal E}\over e_1-e_2}\,=\,4q\,-\,A\,\,,
\label{calE-q-A}
\eeq
where $A$  is given by:
\beq
A={2\over 3}\,(d+1)
\Big[(\gamma-1)^2\,-\,{1\over 2}\,(\alpha-\beta)^2\,-1\,\Big]+
{2-d\over 3}\Big[\,(\epsilon-1)^2\,-\,(\delta-1)^2\,\Big]\,+\,
2\gamma\,[\,\epsilon-d\delta\,]\,\,.\qquad
\eeq
The Heun equation with $d=2$ is a special case. Indeed,  it follows from (\ref{modulus-d}) that the moduli of the elliptic function are
\beq
k\,=\,k'\,=\,{1\over \sqrt{2}}\,\,.
\eeq
Equivalently, one gets from (\ref{tau-modulus}) that the lattice parameter is $\tau=i$ for this case (this is the so-called lemniscatic  lattice). Moreover, see (\ref{es-K}), the $e_i$'s for this case are:
\beq
e_1\,=\,-e_3\,=\,{\big[\,K({1\over \sqrt{2}})\big]^2\over 2\omega_1^2}\,=\,
{\big[\,\Gamma({1\over 4})\big]^2\over 32\pi\omega_1^2}\,\,,\qquad\qquad
e_2\,=\,0\,\,.
\eeq 
Let us write more explicitly the change of variables for this $d=2$ case.
For simplicity we will choose from now on the parameter $\omega_1=1/2$.
 From
(\ref{z-x-general}) we can write $z$ as a function of $x$ in the form:
\beq
z\,=\,1\,-\,{\wp(x)\over e_1}\,=\,
1\,-\,{8\pi \over \big[\,\Gamma({1\over 4})\big]^4}\,\,\wp(x)\,\,.
\label{z-x-d2}
\eeq
The function $\wp(x)$ in (\ref{z-x-d2}) is a doubly periodic function in ${\mathbb C}$ with periods  $(2\omega_1, 2\omega_2)=(1,i)$. When $x$ is on the real line $\wp(x)$ is real, it has a minimum at $x=1/2$ ($\wp(1/2)=e_1$) and $\wp(x)\to+\infty$ at $x=0,1$ . This means that  $z$ runs twice over the interval $[-\infty, 0]$ when $x\in[0,1]$. Notice that, to parametrize the hanging brane configuration of sections \ref{dimers} and \ref{dimer-fluct}, the coordinate $\rho$ defined  in (\ref{rho-barE}) must run twice over the interval $[1,\infty]$, which precisely corresponds to taking  $x\in[0,1]$.

 It is also interesting to write the inverse of the change of variables (\ref{z-x-general}) for the $d=2$ case. By differentiating (\ref{z-x-general}) and making use of (\ref{wp-dif-eq}) for 
$g_2=4\,e_1^2$ and $g_3=0$, one gets:
\beq
{dz\over dx}\,=\,\pm 2\sqrt{e_1}\,\sqrt{z(1-z)(z-2)}\,\,,
\eeq
where the $+$ ($-$) sign must be taken when  $x\in [0,1/2]$ 
($x\in [1/2, 1]$).  This equation can be integrated in the form:
\beq
x\,=\,{1\over 2}\,\Big[\,1\,\pm 
{2\sqrt{2\pi}\over \big[\,\Gamma({1\over 4})\big]^2}\,\,
\int_{0}^z\,{dt\over \sqrt{t(1-t)(t-2)}}\,\Big]\,\,,
\eeq
where we have already taken into account that $x=1/2$ for $z=0$.



\begin{thebibliography}{99}

 \bibitem{jm} J.~M.~Maldacena, ``The large $N$ limit of superconformal field
theories and supergravity'', {\it Adv.\ Theor.\ Math.\ Phys.}\  {\bf 
2} (1998) 231,
{\rm hep-th/9711200}.

 
 
\bibitem{Aharony:1999ti}
  O.~Aharony, S.~S.~Gubser, J.~M.~Maldacena, H.~Ooguri and Y.~Oz,
  ``Large N field theories, string theory and gravity,''
  Phys.\ Rept.\  {\bf 323}, 183 (2000)
  [arXiv:hep-th/9905111].

\bibitem{Pawelczyk:2000hy}
  J.~Pawelczyk and S.~J.~Rey,
  ``Ramond-Ramond flux stabilization of D-branes,''
  Phys.\ Lett.\  B {\bf 493}, 395 (2000)
  [arXiv:hep-th/0007154].



\bibitem{Camino:2001at}
  J.~M.~Camino, A.~Paredes, A.~V.~Ramallo,
  ``Stable wrapped branes,''
  JHEP {\bf 0105}, 011 (2001).
  [hep-th/0104082].



\bibitem{Yamaguchi:2006tq}
  S.~Yamaguchi,
  ``Wilson loops of anti-symmetric representation and D5-branes,''
  JHEP {\bf 0605}, 037 (2006)
  [arXiv:hep-th/0603208].



\bibitem{Gomis:2006sb}
  J.~Gomis and F.~Passerini,
  ``Holographic Wilson loops,''
  JHEP {\bf 0608}, 074 (2006)
  [arXiv:hep-th/0604007].




 
\bibitem{Kachru:2009xf}
  S.~Kachru, A.~Karch and S.~Yaida,
  ``Holographic Lattices, Dimers, and Glasses,''
  Phys.\ Rev.\  D {\bf 81} (2010) 026007
  [arXiv:0909.2639 [hep-th]].


\bibitem{Kachru:2010dk}
  S.~Kachru, A.~Karch and S.~Yaida,
  ``Adventures in Holographic Dimer Models,''
  New J.\ Phys.\  {\bf 13}, 035004 (2011)
  [arXiv:1009.3268 [hep-th]].
  
  
  

\bibitem{Mueck:2010ja}
  W.~Mueck,
  ``The Polyakov Loop of Anti-symmetric Representations as a Quantum Impurity
  Model,''
  Phys.\ Rev.\  D {\bf 83}, 066006 (2011)
  [arXiv:1012.1973 [hep-th]].

\bibitem{Harrison:2011fs}
  S.~Harrison, S.~Kachru and G.~Torroba,
 ``A maximally supersymmetric Kondo model,''
 [ arXiv:1110.5325 [hep-th]].


\bibitem{Faraggi:2011bb}
  A.~Faraggi and L.~A.~Pando Zayas,
  ``The Spectrum of Excitations of Holographic Wilson Loops,''
  JHEP {\bf 1105}, 018 (2011)
  [arXiv:1101.5145 [hep-th]].

\bibitem{Faraggi-dos}
 A.~Faraggi, W. Mueck  and L.~A.~Pando Zayas,
 ``One loop effective action of the holographic antisymmetric Wilson loop", 
 [arXiv:1112.5028 [hep-th]].

 
 



\bibitem{Aharony:2008ug}
  O.~Aharony, O.~Bergman, D.~L.~Jafferis and J. M. Maldacena,
``N=6 superconformal Chern-Simons-matter theories, M2-branes and their gravity duals,'' JHEP {\bf 0810}, 091 (2008).
  [arXiv:0806.1218 [hep-th]].
  



 
\bibitem{BL}
  J.~Bagger, N.~Lambert,
 ``Modeling Multiple M2's,''
  Phys.\ Rev.\  {\bf D75}, 045020 (2007).
  [hep-th/0611108];
  J.~Bagger, N.~Lambert,
 ``Gauge symmetry and supersymmetry of multiple M2-branes,''
  Phys.\ Rev.\  {\bf D77 } (2008)  065008.
  [arXiv:0711.0955 [hep-th]];
  J.~Bagger, N.~Lambert,
 ``Comments on multiple M2-branes,''
  JHEP {\bf 0802}, 105 (2008).
  [arXiv:0712.3738 [hep-th]].
 
\bibitem{Gustavsson:2007vu}
  A.~Gustavsson,
  ``Algebraic structures on parallel M2-branes,''
  Nucl.\ Phys.\  {\bf B811}, 66-76 (2009).
  [arXiv:0709.1260 [hep-th]].
 
 
 
  
  
  
  
  
  
\bibitem{Drukker:2008zx}
  N.~Drukker, J.~Plefka, D.~Young,
  ``Wilson loops in 3-dimensional N=6 supersymmetric Chern-Simons Theory and their string theory duals,''
  JHEP {\bf 0811}, 019 (2008).
  [arXiv:0809.2787 [hep-th]].



\bibitem{Gubser:1998vd}
  S.~S.~Gubser,
  ``Einstein manifolds and conformal field theories,''
  Phys.\ Rev.\  D {\bf 59}, 025006 (1999)
  [arXiv:hep-th/9807164].

\bibitem{Callan:1998iq}
  C.~G.~Callan, A.~Guijosa and K.~G.~Savvidy,
  ``Baryons and string creation from the fivebrane worldvolume action,''
  Nucl.\ Phys.\  B {\bf 547} (1999) 127
  [arXiv:hep-th/9810092].

\bibitem{Callan:1999zf}
  C.~G.~Callan, A.~Guijosa, K.~G.~Savvidy and O.~Tafjord,
  ``Baryons and flux tubes in confining gauge theories from brane actions,''
  Nucl.\ Phys.\  B {\bf 555}, 183 (1999),
  [arXiv:hep-th/9902197].



\bibitem{Skenderis:2002wp}K. Skenderis,  ``Lecture notes on holographic renormalization",   Class. Quant. Grav., {\bf 19} (2002) 5849, 
 [arXiv:hep-th/0209067].


\bibitem{Karch:2005ms}A. Karch,  A.  O'Bannon and K. Skenderis, 
``Holographic renormalization of probe D-branes in AdS/CFT", 
JHEP {\bf 0604}, 015 (2006),  [arXiv:hep-th/0512125].





\bibitem{Maldacena:1998im}
  J.~M.~Maldacena,
  ``Wilson loops in large N field theories,''
  Phys.\ Rev.\ Lett.\  {\bf 80}, 4859 (1998)
  [arXiv:hep-th/9803002].


\bibitem{Rey:1998ik}
  S.~J.~Rey and J.~T.~Yee,
  ``Macroscopic strings as heavy quarks in large N gauge theory and  anti-de
  Sitter supergravity,''
  Eur.\ Phys.\ J.\  C {\bf 22} (2001) 379
  [arXiv:hep-th/9803001].



\bibitem{Klebanov:2006jj}
  I.~R.~Klebanov, J.~M.~Maldacena and C.~B.~Thorn,
  ``Dynamics of flux tubes in large N gauge theories,''
  JHEP {\bf 0604}, 024 (2006)
  [arXiv:hep-th/0602255].



\bibitem{Brower:2006hf}
  R.~C.~Brower, C.~I.~Tan and C.~B.~Thorn,
  ``String / flux tube duality on the lightcone,''
  Phys.\ Rev.\  D {\bf 73}, 124037 (2006)
  [arXiv:hep-th/0603256].



\bibitem{Lame} E. Whittaker and G. Watson, ``A course of modern analysis", Cambridge University Press, 1927; E. Ince, ordinary differential equations, Dover Publications, 1956. 




\bibitem{RS}J. G. Russo and K. Sfetsos,
``Rotating D3-branes and QCD in three dimensions",
{\sl \atmp} {\bf 3}(1999) 131,   [arXiv:hep-th/9901056].


\bibitem{Lin:2004nb}
  H.~Lin, O.~Lunin and J.~M.~Maldacena,
  ``Bubbling AdS space and 1/2 BPS geometries,''
  JHEP {\bf 0410}, 025 (2004)
  [arXiv:hep-th/0409174].



\bibitem{Yamaguchi:2006te}
  S.~Yamaguchi,
  ``Bubbling geometries for half BPS Wilson lines,''
  Int.\ J.\ Mod.\ Phys.\  A {\bf 22}, 1353 (2007)
  [arXiv:hep-th/0601089].



\bibitem{Lunin:2006xr}
  O.~Lunin,
  ``On gravitational description of Wilson lines,''
  JHEP {\bf 0606}, 026 (2006)
  [arXiv:hep-th/0604133].



\bibitem{D'Hoker:2007fq}
  E.~D'Hoker, J.~Estes and M.~Gutperle,
  ``Gravity duals of half-BPS Wilson loops,''
  JHEP {\bf 0706}, 063 (2007)
  [arXiv:0705.1004 [hep-th]].






\bibitem{Karaiskos:2011kf}
  N.~Karaiskos, K.~Sfetsos and E.~Tsatis,
  ``Brane embeddings in sphere submanifolds,''
  arXiv:1106.1200 [hep-th].


\bibitem{Kondopaper}  P. Benincasa and A. V. Ramallo, ``Holographic Kondo model in various dimensions", to appear. 
   
\bibitem{sage}W. A. Stein and others, The Sage Development Team, 
{{S}age {M}athematics {S}oftware ({V}ersion 4.6.1)},
{ http://www.sagemath.org}, {2011}.


 
 
\bibitem{maxima}Maxima, a Computer Algebra System.  Version 5.25.1,
{http://maxima.sourceforge.net/}, {2011}.


\bibitem{scipy} Eric Jones and Travis Oliphant and Pearu Peterson and others, 
 {{SciPy}: Open source scientific tools for {Python}},  {2001--},
 {http://www.scipy.org/}.
 






\bibitem{Craps:1999nc}
  B.~Craps, J.~Gomis, D.~Mateos and A.~Van Proeyen,
  ``BPS solutions of a D5-brane world volume in a D3-brane background from
  superalgebras,''
  JHEP {\bf 9904}, 004 (1999)
  [arXiv:hep-th/9901060].


\bibitem{Camino:1999xx}
  J.~M.~Camino, A.~V.~Ramallo and J.~M.~Sanchez de Santos,
  ``Worldvolume dynamics of D-branes in a D-brane background,''
  Nucl.\ Phys.\  B {\bf 562} (1999) 103
  [arXiv:hep-th/9905118].



\bibitem{Gomis:1999xs}
  J.~Gomis, A.~V.~Ramallo, J.~Simon and P.~K.~Townsend,
  ``Supersymmetric baryonic branes,''
  JHEP {\bf 9911}, 019 (1999),
  [arXiv:hep-th/9907022].





\bibitem{Ceresole:1999ht}
  A.~Ceresole, G.~Dall'Agata and R.~D'Auria,
  ``K K spectroscopy of type IIB supergravity on AdS(5) x T**11,''
  JHEP {\bf 9911} (1999) 009
  [arXiv:hep-th/9907216].
  
  
  
  
\bibitem{Ceresole:1999zs}
  A.~Ceresole, G.~Dall'Agata, R.~D'Auria and S.~Ferrara,
  ``Spectrum of type IIB supergravity on AdS(5) x T**11: Predictions on N=1
  SCFT's,''
  Phys.\ Rev.\  D {\bf 61} (2000) 066001
  [arXiv:hep-th/9905226].



\bibitem{Ronveaux} A. Ronveaux, ``Heun's differential equations", Oxford University Press, 1995. 

 



\bibitem{Smirnov} A. O. Smirnov, ``Elliptic solitons and Heun's equation", 
 [arXiv:math/0109149].


\bibitem{Takemura}
K. Takemura, ``Heun equations and Inozemtsev models",  [arXiv:nlin/0303005];
K. Takemura, ``On eigenvalues of Lam\'e operator", [arXiv:math/0409247];
K. Takemura, ``Finte-gap potential, Heun's differential equation and WKB analysis", RIMS K\^oky\^uroku Bessatsu {\bf B5}, 61-74. 




\bibitem{Maier} R. S. Maier, ``Lam\'e polynomials, hyperelliptic reductions and Lam\'e band structure", Phil. Trans. R. Soc. A2008 {\bf 366}, 1115.


\bibitem{NIST} F. W. J. Olver, D. W. Lozier, R. F. Boisvert and C. W. Clark,
``NIST Handbook of mathematical functions", Cambridge University Press, 2010. 
 
  


    
    
    
    
  
\end{thebibliography}
\end{document}